\documentclass[a4paper,12pt]{article}

\pdfoutput=1

\usepackage{amsmath}
\usepackage{amssymb}
\usepackage{mathrsfs}
\usepackage{bbm}
\usepackage{bm}
\usepackage{graphicx,subfigure}
\usepackage[numbers,sort&compress]{natbib}
\usepackage{array,longtable}
\usepackage{multirow}
\usepackage{slashed}
\usepackage[utf8]{inputenc}

\usepackage[colorlinks,
            linkcolor=blue,
            filecolor=black,
            anchorcolor=black,
            urlcolor=blue,
            citecolor=blue,
            bookmarks=false,
            ]{hyperref}

\numberwithin{equation}{section}

\newlength{\dinwidth}
\newlength{\dinmargin}
\allowdisplaybreaks

\begin{document}

\title{\bf The measurable angular distribution of $\Lambda_b^0 \to \Lambda_c^+ (\to \Lambda^0 \pi^+)\tau^- (\to \pi^- \nu_\tau)\bar{\nu}_\tau$ decay}

\author{
Quan-Yi Hu$^{1,}$\footnote{qyhu@aynu.edu.cn},
Xin-Qiang Li$^{2,}$\footnote{xqli@mail.ccnu.edu.cn},
Ya-Dong Yang$^{2,}$\footnote{yangyd@mail.ccnu.edu.cn}\,
and
Dong-Hui Zheng$^{2,}$\footnote{zhengdh@mails.ccnu.edu.cn}\\[15pt]
\small $^1$School of Physics and Electrical Engineering, Anyang Normal University, Anyang, Henan 455000, China \\
\small $^2$Institute of Particle Physics and Key Laboratory of Quark and Lepton Physics~(MOE), \\
\small Central China Normal University, Wuhan, Hubei 430079, China}

\date{}
\maketitle

\vspace{-0.2cm}

\begin{abstract}
{\noindent}In $\Lambda_b^0 \to \Lambda_c^+ (\to \Lambda^0 \pi^+) \tau^- \bar{\nu}_\tau $ decay, the solid angle of the final-state particle $\tau^-$ cannot be determined precisely since the decay products of the $\tau^-$ include an undetected $\nu_\tau$. Therefore, the angular distribution of this decay cannot be measured. In this work, we construct a {\it measurable} angular distribution by considering the subsequent decay $\tau^- \to \pi^- \nu_\tau$. The full cascade decay is $\Lambda_b^0 \to \Lambda_c^+ (\to \Lambda^0 \pi^+)\tau^- (\to \pi^- \nu_\tau)\bar{\nu}_\tau$. The three-momenta of the final-state particles $\Lambda^0$, $\pi^+$, and $\pi^-$ can be measured. Considering all Lorentz structures of the new physics (NP) effective operators and an unpolarized initial $\Lambda_b$ state, the five-fold differential angular distribution can be expressed in terms of ten angular observables ${\cal K}_i (q^2, E_\pi)$. By integrating over some of the five kinematic parameters, we define a number of observables, such as the $\Lambda_c$ spin polarization $P_{\Lambda_c}(q^2)$ and the forward-backward asymmetry of $\pi^-$ meson $A_{FB}(q^2)$, both of which can be represented by the angular observables $\widehat{{\cal K}}_i (q^2)$. We provide numerical results for the entire set of the angular observables $\widehat{{\cal K}}_i (q^2)$ and $\widehat{{\cal K}}_i$ both within the Standard Model and in some NP scenarios, which are a variety of best-fit solutions in seven different NP hypotheses. We find that the NP which can resolve the anomalies in $\bar{B} \to D^{(*)} \tau^- \bar{\nu}_\tau$ decays has obvious effects on the angular observables $\widehat{{\cal K}}_i (q^2)$, except $\widehat{{\cal K}}_{1ss} (q^2)$ and $\widehat{{\cal K}}_{1cc} (q^2)$.
\end{abstract}

\newpage

\section{Introduction}
\label{sec:introduction}

The anomalous measurements~\cite{Lees:2012xj,Lees:2013uzd,Aaij:2015yra,Huschle:2015rga,Hirose:2016wfn,Aaij:2017uff,Hirose:2017dxl,Aaij:2017deq,Belle:2019rba} on $\bar{B} \to D^{(*)} \tau^- \bar{\nu}_\tau$ decays indicate the existence of new physics (NP) that breaks the universality of lepton flavour in $b \to c \tau^- \bar{\nu}_\tau$ transition. At the typical energy scale $\mu \simeq m_b$, the $b$-hadron decays involving $b \to c \tau^- \bar{\nu}_\tau$ transition are governed by the following effective Hamiltonian\footnote{In this work, we only consider left-handed neutrinos. The effective Hamiltonian containing right-handed neutrinos can be found in Refs.~\cite{Dutta:2013qaa,Ligeti:2016npd,Mandal:2020htr}. It can be derived from the identity $\sigma^{\mu\nu}\gamma_5 = -\frac{i}{2}\epsilon^{\mu\nu\alpha\beta} \sigma_{\alpha\beta}$ that the operator $({\bar c}\sigma^{\mu\nu}(1+\gamma_5) b)({\bar\tau}\sigma_{\mu\nu} \nu_{\tau L}) $ is absent. We use the convention $\epsilon_{0123} = - \epsilon^{0123} = 1$.} 
\begin{align}
\label{eq:Heff}
{\cal H}_{\rm eff} 
= \sqrt{2} G_F V_{cb} \big[  
& g_V ({\bar c}\gamma^\mu b) ({\bar\tau}\gamma_\mu \nu_{\tau L}) + g_A ({\bar c}\gamma^\mu \gamma_5 b)({\bar\tau}\gamma_\mu \nu_{\tau L}) 
\nonumber\\[2mm]
&+ g_S ({\bar c} b)({\bar\tau} \nu_{\tau L}) + g_P ({\bar c} \gamma_5 b)({\bar\tau} \nu_{\tau L}) 
\nonumber\\[2mm]
&+ g_T ({\bar c}\sigma^{\mu\nu}(1-\gamma_5) b)({\bar\tau}\sigma_{\mu\nu} \nu_{\tau L}) \big] + {\rm H.c.},
\end{align}
where $\sigma^{\mu\nu} = \frac{i}{2}[\gamma^\mu,\, \gamma^\nu]$. $\nu_{\tau L} = P_L \nu_\tau$ is the field of left-handed neutrino. The left-handed chirality projector $P_L = (1-\gamma_5)/2$. In the Standard Model (SM), the Wilson coefficients satisfy $g_V = - g_A = 1$ and $g_S = g_P = g_T = 0$. To understand the anomalies in $\bar{B} \to D^{(*)} \tau^- \bar{\nu}_\tau$ decays, a number of global analyses have been carried out~\cite{Alok:2017qsi,Hu:2018veh,Alok:2019uqc,Murgui:2019czp,Blanke:2018yud,Blanke:2019qrx,Cheung:2020sbq,Kumbhakar:2020jdz}, finding that some different combinations of Wilson coefficients can well explain these anomalies. In addition, a large number of studies have been done in some specific NP models, such as leptoquarks, $R$-parity violating supersymetric models, charged Higgses, and charged vector bosons; see for instance Refs.~\cite{Sakaki:2013bfa,Bauer:2015knc,Fajfer:2015ycq,Crivellin:2017zlb,Iguro:2018vqb,Li:2016vvp,Becirevic:2018afm,Angelescu:2018tyl,Crivellin:2019dwb,Crivellin:2012ye,Celis:2012dk,Deshpand:2016cpw,Altmannshofer:2017poe,Hu:2018lmk,Hu:2020yvs,Ko:2012sv,Celis:2016azn,Boucenna:2016wpr,Datta:2012qk,Altmannshofer:2020axr,Freytsis:2015qca,Greljo:2018ogz,Azatov:2018knx,Blanke:2018sro,Bansal:2018nwp}. In these NP scenarios, the $\Lambda_b^0 \to \Lambda_c^+ \tau^- \bar{\nu}_\tau$ decay, which is also governed by the $b \to c \tau^- \bar{\nu}_\tau$ transition, will receive contributions from the NP.

The baryonic decay $\Lambda_b^0 \to \Lambda_c^+ \tau^- \bar{\nu}_\tau$ could be useful to confirm possible Lorentz structures of the NP effective operators and to distinguish the specific NP models. Due to the spin-half nature of $\Lambda_b$ and $\Lambda_c$ baryons, all the effective operators in Eq.~\eqref{eq:Heff} can affect $\Lambda_b^0 \to \Lambda_c^+ \tau^- \bar{\nu}_\tau$ decay. But for the mesonic counterparts, the operators $({\bar c}\gamma^\mu \gamma_5 b)({\bar\tau}\gamma_\mu \nu_{\tau L})$ and $({\bar c} \gamma_5 b)({\bar\tau} \nu_{\tau L})$ cannot affect the $\bar{B} \to D$ processes, and operator $({\bar c} b)({\bar\tau} \nu_{\tau L})$ cannot affect the $\bar{B} \to D^*$ processes. The large production cross section of $\Lambda_b^0$ on the LHC and the clear $\Lambda_b^0 \to \Lambda_c^+$ transition form factors~\cite{Detmold:2015aaa,Datta:2017aue,Azizi:2018axf,Aaij:2017svr,Bernlochner:2018kxh,Bernlochner:2018bfn} make $\Lambda_b^0 \to \Lambda_c^+ \tau^- \bar{\nu}_\tau$ decay a good candidate to complement the $\bar{B} \to D^{(*)} \tau^- \bar{\nu}_\tau$ decays. In the previous studies of the NP contributions in $\Lambda_b^0 \to \Lambda_c^+ \tau^- \bar{\nu}_\tau$ decay~\cite{Dutta:2015ueb,Li:2016pdv,Hu:2018veh,DiSalvo:2018ngq,Ray:2018hrx,Penalva:2019rgt,Mu:2019bin}, especially in some studies considering the angular distribution of the cascade decay $\Lambda_b^0 \to \Lambda_c^+ (\to \Lambda^0 \pi^+) \tau^- \bar{\nu}_\tau $~\cite{Gutsche:2015mxa,Shivashankara:2015cta,Boer:2019zmp,Ferrillo:2019owd}, the information of the polar and azimuthal angles $(\theta_\tau,\,\phi_\tau)$ of the final-state particle $\tau^-$ may be used. However, as pointed out in Ref.~\cite{Bhattacharya:2020lfm}, the polar and azimuthal angles $(\theta_\tau,\,\phi_\tau)$ cannot be determined precisely since the decay products of the $\tau^-$ include an undetected $\nu_\tau$. Therefore, in this work, we construct a {\it measurable} angular distribution by considering the subsequent decay $\tau^- \to \pi^- \nu_\tau$. The full cascade decay is $\Lambda_b^0 \to \Lambda_c^+ (\to \Lambda^0 \pi^+)\tau^- (\to \pi^- \nu_\tau)\bar{\nu}_\tau$, which includes two undetected final-state particles $\nu_\tau$ and $\bar{\nu}_\tau$, as well as three visible final-state particles whose three momenta can be measured: $\Lambda^0$, $\pi^+$, and $\pi^-$.

Our paper is organized as follows. In Section~\ref{sec:analytical}, we define the independent transversity amplitudes and give the analytical results of the {\it measurable} angular distribution of the five-body decay $\Lambda_b^0 \to \Lambda_c^+ (\to \Lambda^0 \pi^+)\tau^- (\to \pi^- \nu_\tau)\bar{\nu}_\tau$, with an unpolarized $\Lambda_b^0$. Discussions of the integrated observables are included in Section~\ref{sec:observables}. The numerical analyses and results are shown in Section~\ref{sec:numerical}. Our conclusions are finally made in Section~\ref{sec:conclusion}. In the Appendix~\ref{sec:app}, we present the detailed calculation procedures and some conventions.
 
\section{Analytical results}
\label{sec:analytical}

In this section, we directly list the analytical results of the angular distribution. The detailed calculation procedures are presented in the Appendix~\ref{sec:app}.

\subsection{Transversity amplitudes}
\label{subsec:TA}

In order to get the compact form of the analytical results, we adopt the helicity-based definition of the $\Lambda_b\to\Lambda_c$ form factors~\cite{Feldmann:2011xf}, which are given in Ref.~\cite{Datta:2017aue}. The matrix elements of vector and axial vector currents can be expressed by six helicity form factors $F_+$, $F_\perp$, $F_0$, $G_+$, $G_\perp$, and $G_0$. Using the Ward-like identity for the $\Lambda_b\to\Lambda_c$ matrix elements
\begin{equation}
\label{eq:Ward}
\left\langle \Lambda_c\left| \bar{c} \left( \gamma_5 \right) b\right| \Lambda_b \right\rangle 
= \frac{q_\mu}{m_b\mp m_c} \left\langle \Lambda_c\left| \bar{c}  \left( \gamma_5 \right) \gamma^\mu b\right| \Lambda_b \right\rangle,
\end{equation}
the matrix elements of scalar and pseudoscalar currents can be written in terms of $F_0$ and $G_0$, respectively. In the absence of the tensor operator, we can define six independent transversity amplitudes as follows
\begin{align}
&{\cal A}_{\perp_t} = {\cal A}^{SP}_{\perp_t} + \frac{m_\tau}{\sqrt{q^2}} {\cal A}^{VA}_{\perp_t},
\\[2mm]
&{\cal A}_{\parallel_t} = {\cal A}^{SP}_{\parallel_t} + \frac{m_\tau}{\sqrt{q^2}} {\cal A}^{VA}_{\parallel_t},
\\[2mm]
& {\cal A}_{\perp_1} = -2 F_\perp \sqrt{Q_-} g_V,
\\[2mm]
& {\cal A}_{\parallel_1} = -2 G_\perp \sqrt{Q_+} g_A,
\\[2mm]
& {\cal A}_{\perp_0} = F_+ \sqrt{2 Q_-} \frac{m_{\Lambda_b} + m_{\Lambda_c}}{\sqrt{q^2}} g_V,
\\[2mm]
& {\cal A}_{\parallel_0} = G_+ \sqrt{2 Q_+} \frac{m_{\Lambda_b} - m_{\Lambda_c}}{\sqrt{q^2}} g_A.
\end{align}
Here, $\perp$ and $\parallel$ stand for the different transversity states. The subscript $t$ represents time-like $\tau^-\bar{\nu}_\tau$ state; the subscripts $1$ and $0$ denote the magnitude of the $z$-component of the $\tau^-\bar{\nu}_\tau$ angular momentum in the vector $\tau^-\bar{\nu}_\tau$ state. $Q_\pm \equiv (m_{\Lambda_b}\pm m_{\Lambda_c})^2 - q^2$. The time-like transversity amplitudes ${\cal A}^{SP}_{\perp_t}$, ${\cal A}^{SP}_{\parallel_t}$, ${\cal A}^{VA}_{\perp_t}$, and ${\cal A}^{VA}_{\parallel_t}$ are respectively defined as
\begin{align}
&{\cal A}^{SP}_{\perp_t} = F_0 \sqrt{2 Q_+} \frac{m_{\Lambda_b} - m_{\Lambda_c}}{m_b - m_c} g_S,
&&{\cal A}^{SP}_{\parallel_t} = - G_0 \sqrt{2 Q_-} \frac{m_{\Lambda_b} + m_{\Lambda_c}}{m_b + m_c} g_P,
\\[2mm]
&{\cal A}^{VA}_{\perp_t} = F_0 \sqrt{2 Q_+} \frac{m_{\Lambda_b} - m_{\Lambda_c}}{\sqrt{q^2}} g_V,
&&{\cal A}^{VA}_{\parallel_t} = G_0 \sqrt{2 Q_-} \frac{m_{\Lambda_b} + m_{\Lambda_c}}{\sqrt{q^2}} g_A.
\end{align}

The matrix elements of the tensor currents can be expressed by four helicity form factors $h_+$, $h_\perp$, $\widetilde{h}_+$, and $\widetilde{h}_\perp$, and we need to define four additional independent transversity amplitudes as follows
\begin{align}
& {\cal A}^T_{\perp_1} = 4 h_\perp \sqrt{Q_-} \frac{m_{\Lambda_b} + m_{\Lambda_c}}{\sqrt{q^2}} g_T,
\\[2mm]
& {\cal A}^T_{\parallel_1} = 4 \widetilde{h}_\perp \sqrt{Q_+} \frac{m_{\Lambda_b} - m_{\Lambda_c}}{\sqrt{q^2}} g_T,
\\[2mm]
& {\cal A}^T_{\perp_0} = -2 h_+ \sqrt{2 Q_-} g_T,
\\[2mm]
& {\cal A}^T_{\parallel_0} = -2 \widetilde{h}_+ \sqrt{2 Q_+} g_T.
\end{align}
The superscript $T$ indicates that an amplitude arises only when there are tensor operators.

%%%%%%%%%%%%%%%%%%%%%%%%%%%%%%%%%%%%%%%%%%%%
\begin{table}
\begin{center}
\begin{tabular}{|c|c|} 
\hline
Transversity Amplitudes & Couplings
\\ \hline
${\cal A}^{VA}_{\perp_t}$, ${\cal A}_{\perp_1}$, ${\cal A}_{\perp_0}$
& $g_V$
\\
${\cal A}^{VA}_{\parallel_t}$, ${\cal A}_{\parallel_1}$, ${\cal A}_{\parallel_0}$
& $g_A$
\\
${\cal A}^{SP}_{\perp_t}$ & $g_S$
\\
${\cal A}^{SP}_{\parallel_t}$ & $g_P$
\\
${\cal A}_{\perp_t}$ & $g_V$, $g_S$
\\
${\cal A}_{\parallel_t}$ & $g_A$, $g_P$
\\
${\cal A}^T_{\perp_1}$, ${\cal A}^T_{\parallel_1}$, ${\cal A}^T_{\perp_0}$, ${\cal A}^T_{\parallel_0}$ 
& $g_T$
\\ \hline
\end{tabular}
\caption{\label{tab:NPTA} \small Contributions of the NP Wilson coefficients to the various transversity amplitudes.}
\end{center}
\end{table}
%%%%%%%%%%%%%%%%%%%%%%%%%%%%%%%%%%%%%%%%%%%%

\subsection{Angular distribution}
\label{subsec:angular}

%%%%%%%%%%%%%%%%%%%%%%%%%%%%%%%%%
\begin{figure}[t]
\centering
\includegraphics[width=0.45\textwidth]{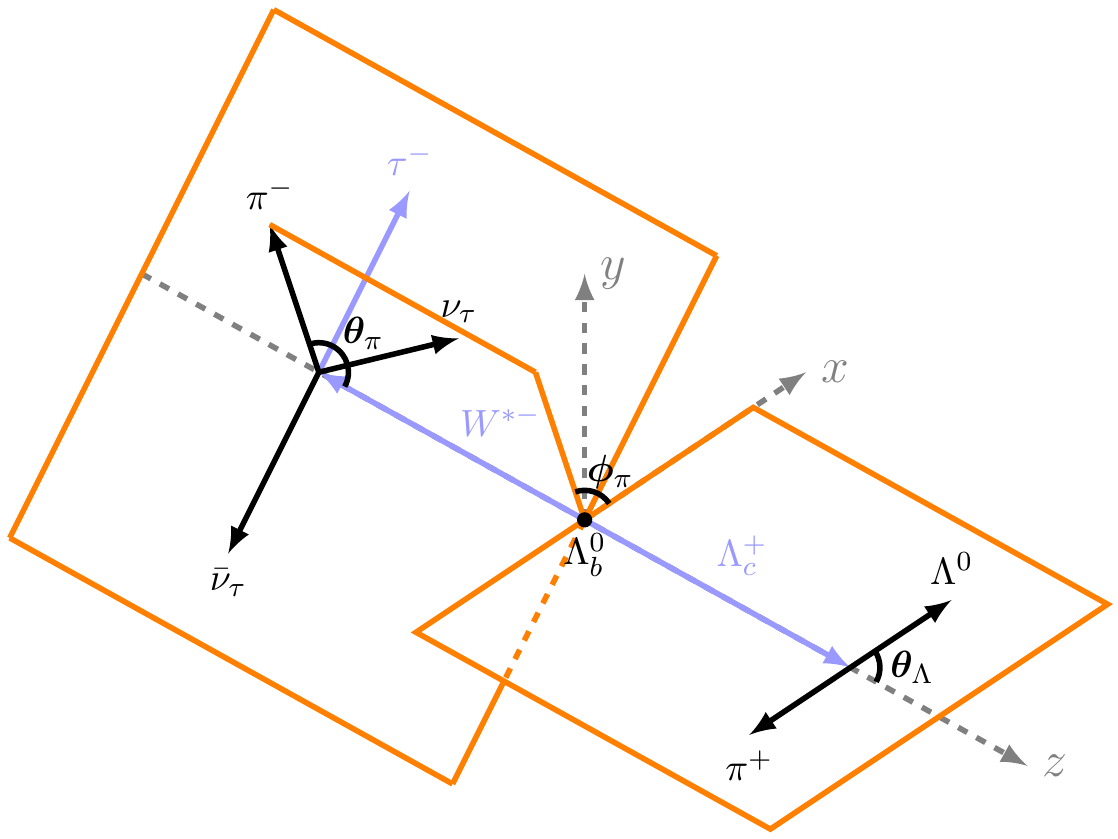}
\caption{\label{fig:angle} \small Definition of the angles in the unpolarized $\Lambda_b^0 \to \Lambda_c^+ (\to \Lambda^0 \pi^+)\tau^- (\to \pi^- \nu_\tau)\bar{\nu}_\tau$ decay.}
\end{figure}
%%%%%%%%%%%%%%%%%%%%%%%%%%%%%%%%%

The {\it measurable} angular distribution of the five-body $\Lambda_b^0 \to \Lambda_c^+ (\to \Lambda^0 \pi^+)\tau^- (\to \pi^- \nu_\tau)\bar{\nu}_\tau$ decay, with an unpolarized $\Lambda_b^0$, is described by the $\tau^-\bar{\nu}_\tau$ invariant mass squared $q^2$; the helicity angle of $\Lambda^0$ baryon in the $\Lambda_c^+$ rest frame, $\theta_\Lambda$; as well as the energy, the polar angle, and the azimuthal angle of $\pi^-$ in the $\tau^-\bar{\nu}_\tau$ center-of-mass frame, $E_\pi$, $\theta_\pi$, and $\phi_\pi$. For more details, we refer to Figure~\ref{fig:angle} and the Appendix~\ref{sec:app}. The five-fold differential decay rate can then be written as 
\begin{align}
\frac{d^5\Gamma}{dq^2 dE_\pi d\cos\theta_\pi d\phi_\pi d\cos\theta_\Lambda}
=& \frac{G_F^2 \left| V_{cb}\right|^2  \left| {\bm p}_{\Lambda_c}\right| (q^2)^{3/2} m_\tau^2}{256 \pi^4 m_{\Lambda_b}^2 (m_\tau^2 - m_\pi^2)^2} {\cal B}(\tau \to \pi^- \nu_\tau) {\cal B}(\Lambda_c \to \Lambda \pi^+)
\nonumber \\[2mm]
& \times {\cal K}\left(q^2,E_\pi,\cos\theta_{\Lambda}, \cos\theta_\pi, \phi_\pi \right),  
\label{eq:5fold}
\end{align}
where $\left| {\bm p}_{\Lambda_c}\right| = \sqrt{Q_+ Q_-}/(2 m_{\Lambda_b})$ is the magnitude of the $\Lambda_c$ three-momentum in the $\Lambda_b$ rest frame. By rearranging ${\cal N}^S_i |{\cal A}_i|^2$, ${\cal N}^R_{i,j} {\rm Re}[{\cal A}_i {\cal A}_j^*]$, and ${\cal N}^I_{i,j} {\rm Im}[{\cal A}_i {\cal A}_j^*]$ pieces of Eq.~\eqref{eq:FFDW}, which are listed in Table~\ref{tab:NS}, \ref{tab:NR}, and \ref{tab:NI}, the angular distribution ${\cal K}$ can be expressed as a set of trigonometric functions as follows
\begin{align}
{\cal K}\left(q^2,E_\pi,\cos\theta_{\Lambda}, \cos\theta_\pi, \phi_\pi \right)
=&\sum_{i=1}^{10} {\cal K}_i (q^2, E_\pi) \Omega_i(\cos\theta_{\Lambda}, \cos\theta_\pi, \phi_\pi)
\nonumber \\[2mm]
\equiv &\left({\cal K}_{1ss} \sin^2\theta_\pi + {\cal K}_{1cc} \cos^2\theta_\pi + {\cal K}_{1c} \cos\theta_\pi \right) 
\nonumber \\[2mm]
&+ \left({\cal K}_{2ss} \sin^2\theta_\pi + {\cal K}_{2cc} \cos^2\theta_\pi + {\cal K}_{2c} \cos\theta_\pi \right) \cos\theta_{\Lambda}
\nonumber \\[2mm]
&+ \left({\cal K}_{3sc} \sin\theta_\pi \cos\theta_\pi + {\cal K}_{3s} \sin\theta_\pi \right) \sin\theta_{\Lambda} \sin\phi_\pi
\nonumber \\[2mm]
&+ \left({\cal K}_{4sc} \sin\theta_\pi \cos\theta_\pi + {\cal K}_{4s} \sin\theta_\pi \right) \sin\theta_{\Lambda} \cos\phi_\pi, \label{eq:calK}
\end{align}
where the ten angular observables ${\cal K}_i (q^2, E_\pi)$  can be completely expressed by the transversity amplitudes, the dimensionless factors (see Eqs.~\eqref{eq:St}--\eqref{eq:R3}), and the asymmetry parameter $\alpha_{\Lambda_c}$ (see Eq.~\eqref{eq:alphaLambdac}) as follows 
\begin{align}
{\cal K}_{1ss} =
& \Big[ S_t \left|{\cal A}_{\perp_t}\right|^2 + \left(S_1-S_3\right) \left|{\cal A}_{\perp_1}\right|^2 + \left(S_1+S_3\right) \left|{\cal A}_{\perp_0}\right|^2 
\nonumber \\[2mm]
&+ \left(S_1^T-S_3^T\right) \left|{\cal A}^T_{\perp_1}\right|^2 + \left(S_1^T+S_3^T\right) \left|{\cal A}^T_{\perp_0}\right|^2 + \left(\perp \leftrightarrow \parallel \right) \Big] 
\nonumber \\[2mm]
& + {\rm Re} \left[\left(R_1-R_3\right) {\cal A}_{\perp_1} {\cal A}^{T*}_{\perp_1} + \left(R_1+R_3\right) {\cal A}_{\perp_0} {\cal A}^{T*}_{\perp_0} + \left(\perp \leftrightarrow \parallel \right) \right], 
\label{eq:K1ss}
\\[2mm]
{\cal K}_{1cc} =
&\Big[ S_t \left|{\cal A}_{\perp_t}\right|^2 + \left(S_1+S_3\right) \left|{\cal A}_{\perp_1}\right|^2 + \left(S_1-3S_3\right) \left|{\cal A}_{\perp_0}\right|^2 
\nonumber \\[2mm]
&+ \left(S_1^T+S_3^T\right) \left|{\cal A}^T_{\perp_1}\right|^2 + \left(S_1^T-3S_3^T\right) \left|{\cal A}^T_{\perp_0}\right|^2 + \left(\perp \leftrightarrow \parallel \right) \Big] 
\nonumber \\[2mm] 
&+ {\rm Re} \left[\left(R_1+R_3\right) {\cal A}_{\perp_1} {\cal A}^{T*}_{\perp_1} + \left(R_1-3R_3\right) {\cal A}_{\perp_0} {\cal A}^{T*}_{\perp_0} + \left(\perp \leftrightarrow \parallel \right) \right], 
\label{eq:K1cc}
\\[2mm] 
{\cal K}_{1c} =
& 2{\rm Re} \left[S_2 {\cal A}_{\perp_1} {\cal A}^{*}_{\parallel_1} + S_2^T {\cal A}^T_{\perp_1} {\cal A}^{T*}_{\parallel_1}\right] 
\nonumber \\[2mm]
&+ {\rm Re}\left[ R_2{\cal A}_{\perp_1} {\cal A}^{T*}_{\parallel_1} - \sqrt{2} R_t {\cal A}_{\perp_t} {\cal A}^{*}_{\perp_0} - \sqrt{2} R_t^T {\cal A}_{\perp_t} {\cal A}^{T*}_{\perp_0} + \left( \perp \leftrightarrow \parallel\right) \right], 
\label{eq:K1c}
\\[2mm] 
{\cal K}_{2ss} = 
&2 \alpha_{\Lambda_c} {\rm Re}\Big[S_t {\cal A}_{\perp_t} {\cal A}^{*}_{\parallel_t} + \left(S_1-S_3\right) {\cal A}_{\perp_1} {\cal A}^{*}_{\parallel_1} + \left(S_1+S_3\right) {\cal A}_{\perp_0} {\cal A}^{*}_{\parallel_0} 
\nonumber \\[2mm] 
&+ \left(S_1^T-S_3^T\right) {\cal A}^T_{\perp_1} {\cal A}^{T*}_{\parallel_1} + \left(S_1^T+S_3^T\right) {\cal A}^T_{\perp_0} {\cal A}^{T*}_{\parallel_0} \Big] 
\nonumber \\[2mm] 
& + \alpha_{\Lambda_c} {\rm Re}\left[\left(R_1+R_3\right){\cal A}_{\perp_0} {\cal A}^{T*}_{\parallel_0} + \left(R_1-R_3\right){\cal A}_{\perp_1} {\cal A}^{T*}_{\parallel_1} + \left( \perp \leftrightarrow \parallel\right) \right], 
\label{eq:K2ss}
\\[2mm] 
{\cal K}_{2cc} = 
&2 \alpha_{\Lambda_c} {\rm Re}\Big[S_t {\cal A}_{\perp_t} {\cal A}^{*}_{\parallel_t} + \left(S_1+S_3\right) {\cal A}_{\perp_1} {\cal A}^{*}_{\parallel_1} + \left(S_1-3S_3\right) {\cal A}_{\perp_0} {\cal A}^{*}_{\parallel_0} 
\nonumber \\[2mm]
&+ \left(S_1^T+S_3^T\right) {\cal A}^T_{\perp_1} {\cal A}^{T*}_{\parallel_1} + \left(S_1^T-3S_3^T\right) {\cal A}^T_{\perp_0} {\cal A}^{T*}_{\parallel_0} \Big] 
\nonumber \\[2mm] 
&+ \alpha_{\Lambda_c} {\rm Re}\left[\left(R_1+R_3\right) {\cal A}_{\perp_1} {\cal A}^{T*}_{\parallel_1} + \left(R_1-3R_3\right) {\cal A}_{\perp_0} {\cal A}^{T*}_{\parallel_0} + \left( \perp \leftrightarrow \parallel\right) \right], 
\label{eq:K2cc}
\\[2mm]
{\cal K}_{2c} = 
&\alpha_{\Lambda_c} \left[S_2 \left| {\cal A}_{\perp_1}\right|^2 + S_2^T \left| {\cal A}^T_{\perp_1}\right|^2 + \left( \perp \leftrightarrow \parallel\right) \right] 
\nonumber \\[2mm] 
&+\alpha_{\Lambda_c} {\rm Re}\left[R_2 {\cal A}_{\perp_1} {\cal A}^{T*}_{\perp_1} -\sqrt{2} R_t {\cal A}_{\perp_t} {\cal A}^{*}_{\parallel_0} -\sqrt{2} R_t^T {\cal A}_{\perp_t} {\cal A}^{T*}_{\parallel_0} + \left( \perp \leftrightarrow \parallel\right)\right], 
\label{eq:K2c}
\\[2mm]
{\cal K}_{3sc} = 
&2\sqrt{2}\alpha_{\Lambda_c} {\rm Im}\Big[2S_3 {\cal A}_{\perp_1} {\cal A}^{*}_{\perp_0} + 2S_3^T {\cal A}^T_{\perp_1} {\cal A}^{T*}_{\perp_0}
\nonumber \\[2mm]
&+ R_3 {\cal A}_{\perp_1} {\cal A}^{T*}_{\perp_0} - R_3 {\cal A}_{\perp_0} {\cal A}^{T*}_{\perp_1} - \left( \perp \leftrightarrow \parallel\right) \Big], 
\label{eq:K3sc}
\\[2mm]
{\cal K}_{3s} = 
&-\frac{\alpha_{\Lambda_c}}{\sqrt{2}} {\rm Im}\Big[\sqrt{2}R_t {\cal A}_{\perp_t} {\cal A}^{*}_{\perp_1} + \sqrt{2}R_t^T {\cal A}_{\perp_t} {\cal A}^{T*}_{\perp_1} + 2S_2 {\cal A}_{\perp_1} {\cal A}^{*}_{\parallel_0} 
\nonumber \\[2mm]
&+ R_2 {\cal A}_{\perp_1} {\cal A}^{T*}_{\parallel_0} + R_2 {\cal A}_{\perp_0} {\cal A}^{T*}_{\parallel_1} + 2S_2^T {\cal A}^T_{\perp_1} {\cal A}^{T*}_{\parallel_0} - \left( \perp \leftrightarrow \parallel\right)\Big],\label{eq:K3s}
\\[2mm]
{\cal K}_{4sc} = 
&2\sqrt{2}\alpha_{\Lambda_c} {\rm Re}\Big[R_3 {\cal A}_{\perp_0} {\cal A}^{T*}_{\parallel_1} -2S_3 {\cal A}_{\perp_1} {\cal A}^{*}_{\parallel_0} 
\nonumber \\[2mm]
&- R_3 {\cal A}_{\perp_1} {\cal A}^{T*}_{\parallel_0} -2S_3^T {\cal A}^T_{\perp_1} {\cal A}^{T*}_{\parallel_0} - \left( \perp \leftrightarrow \parallel\right) \Big], 
\label{eq:K4sc}
\\[2mm]
{\cal K}_{4s} = 
&\frac{\alpha_{\Lambda_c}}{\sqrt{2}} {\rm Re}\Big[\sqrt{2}R_t {\cal A}_{\perp_t} {\cal A}^{*}_{\parallel_1} + \sqrt{2}R_t^T {\cal A}_{\perp_t} {\cal A}^{T*}_{\parallel_1} + 2S_2 {\cal A}_{\perp_1} {\cal A}^{*}_{\perp_0} 
\nonumber \\[2mm]
&+ R_2 {\cal A}_{\perp_1} {\cal A}^{T*}_{\perp_0} + R_2 {\cal A}_{\perp_0} {\cal A}^{T*}_{\perp_1} + 2S_2^T {\cal A}^T_{\perp_1} {\cal A}^{T*}_{\perp_0} - \left( \perp \leftrightarrow \parallel\right) \Big]. 
\label{eq:K4s}
\end{align}

The time-like pieces of our results can be completely formulated by transversity amplitudes ${\cal A}_{\perp_t}$ and ${\cal A}_{\parallel_t}$, without using ${\cal A}^{SP}_{\perp_t}$, ${\cal A}^{SP}_{\parallel_t}$, ${\cal A}^{VA}_{\perp_t}$, and/or ${\cal A}^{VA}_{\parallel_t}$. This is consistent with the Ward-like relation~\eqref{eq:Ward}. We can obtain the differential decay rate $d\Gamma/dq^2$ as a function of $q^2$ by integrating over $E_\pi$, $\cos\theta_{\Lambda}$, $\cos\theta_\pi$, and $\phi_\pi$. Apart from the factors ${\cal B}(\tau \to \pi^- \nu_\tau)$ and ${\cal B}(\Lambda_c \to \Lambda \pi^+)$, we find that our results of $d\Gamma/dq^2$ (see Eq.~\eqref{eq:BRq2}) are complete agreement with those in Ref.~\cite{Datta:2017aue}, which discusses the $\Lambda_b \to \Lambda_c \tau \bar{\nu}_\tau$ decay in the presence of all dimension-six operators.

In the SM, $g_V = - g_A = 1$ and $g_S = g_P = g_T = 0$, the angular observables ${\cal K}_{3sc}$ and ${\cal K}_{3s}$ are vanishing. Therefore, a non-vanishing ${\cal K}_{3sc}$ or ${\cal K}_{3s}$ indicates that there is NP effect, which induces a complex contribution to the amplitude. 
\begin{itemize}
\item Suppose that the angular distribution is found to contain the component $\sin\theta_\pi \cos\theta_\pi \sin\theta_\Lambda \sin\phi_\pi$. This indicates that at least one of ${\rm Im}[{\cal A}_{\perp_1} {\cal A}^{T*}_{\perp_0}]$, ${\rm Im}[{\cal A}_{\perp_0} {\cal A}^{T*}_{\perp_1}]$, ${\rm Im}[{\cal A}_{\parallel_1} {\cal A}^{T*}_{\parallel_0}]$, and ${\rm Im}[{\cal A}_{\parallel_0} {\cal A}^{T*}_{\parallel_1}]$ is nonzero, which implies that $g_T \neq 0$, and that the $g_T$ has a different phase than $g_V$ or $g_A$.
\item Suppose that the angular distribution is found to contain the component $\sin\theta_\pi \sin\theta_\Lambda \sin\phi_\pi$. This indicates that the imaginary part of at least one of $g_S g_V^*$, $g_P g_A^*$, $g_V g_A^*$, $g_S g_T^*$, $g_P g_T^*$, $g_V g_T^*$, and $g_A g_T^*$ is not equal to zero.
\end{itemize} 

Here we give a direct way to determine the existence of the tensor operator, that is, a nonzero angular observable ${\cal K}_{3sc}$ would be a solid signal of tensor-type NP. In fact, the angular distribution ${\cal K}$ can also be written in terms of the Wigner $D$-functions, and the terms corresponding to the last two lines of Eq.~\eqref{eq:calK} are given by $2{\rm Re}\left[{\cal K}^{1,1}_1 \Omega^{1,1}_1 \left(\Omega_\Lambda, \Omega_\pi\right) + {\cal K}^{1,2}_1 \Omega^{1,2}_1 \left(\Omega_\Lambda, \Omega_\pi\right)\right]$~\cite{Gratrex:2015hna}, where $\Omega_\Lambda = \left(0, \theta_\Lambda, 0\right)$ and $\Omega_\pi = \left(\phi_\pi, \theta_\pi, -\phi_\pi\right)$. The ${\cal K}^{l_\Lambda,l_\pi}_m$ are given by
\begin{equation}
{\cal K}^{1,2}_1 = \frac{1}{\sqrt{3}} \left({\cal K}_{4sc} + i {\cal K}_{3sc} \right),
\quad
{\cal K}^{1,1}_1 = {\cal K}_{4s} + i {\cal K}_{3s},
\end{equation}
and the $\Omega^{l_\Lambda,l_\pi}_m\left(\Omega_\Lambda, \Omega_\pi\right) = D^{l_\Lambda}_{m,0}\left(\Omega_\Lambda\right) D^{l_\pi}_{m,0}\left(\Omega_\pi\right)$, with the explicit Wigner $D$-functions used here are given by
\begin{equation}
D^1_{1,0}\left(\phi, \theta, -\phi\right) = -\frac{1}{\sqrt{2}} \sin\theta \,e^{-i \phi},
\quad
D^2_{1,0}\left(\phi, \theta, -\phi\right) = -\sqrt{\frac{3}{8}} \sin2\theta \,e^{-i \phi}.
\end{equation}
Each Wigner $D$-function in $\Omega^{l_\Lambda,l_\pi}_m$ is derived by reducing the pair of Wigner $D$-functions generated by the squared matrix element to single one by the Clebsch-Gordan series. So the range of the indices $l_\Lambda$ and $l_\pi$ is $0\leq l_{\Lambda (\pi)} \leq 2 \max \left[J_{\Lambda_c (W)}\right]$. The $J_{\Lambda_c} = \frac{1}{2}$ leads to $0\leq l_\Lambda \leq 1$. In the lepton-side factorization approximation, only $J_W\leq 1$ contributions come from the dimension-six operators in effective Hamiltonian ${\cal H}_{\rm eff}$, thus resulting in $0\leq l_\pi \leq 2$. There is no $J_W = 2$ partial wave  since the two indices in the tensor operator are antisymmetric and therefore in a spin-1 representation. The time-like contributions, which are induced by scalar and pseudo-scalar operators as well as the spin-0 components of the vector and axial-vector operators, are not included in the factor ${\cal K}^{1,2}_1$ (or the angular observables ${\cal K}_{3sc}$ and ${\cal K}_{4sc}$), because they need to be combined with the contribution of an operator in spin-2 representation to produce $l_\pi=2$. Furthermore, there is no imaginary part in ${\cal A}_{\perp_1} {\cal A}^{*}_{\perp_0} \sim \left|g_V\right|^2$ and ${\cal A}_{\parallel_1} {\cal A}^{*}_{\parallel_0} \sim \left|g_A\right|^2$. These make it possible to directly determine the existence of the NP tensor operators by using a nonzero angular observable ${\cal K}_{3sc}$. The similar direct conclusion does not exist for ${\cal K}_{3s}$ since $l_\pi=1$ can be produced by various combinations, each of which contains contribution from at least one operator in spin-1 representation. Therefore, the Eq.~\eqref{eq:K3s} does not contain terms that are only combined by time-like transversity amplitudes.

\section{Observables}
\label{sec:observables}

The five-fold differential decay rate of $\Lambda_b^0 \to \Lambda_c^+ (\to \Lambda^0 \pi^+)\tau^- (\to \pi^- \nu_\tau)\bar{\nu}_\tau$ decay depends on five {\it measurable} kinematic parameters $q^2$, $E_\pi$, $\theta_{\Lambda}$, $\theta_\pi$, and $\phi_\pi$, and a complete experimental analysis may be limited by statistics. By integrating some kinematic parameters, abundant observables can be constructed.

\subsection{Hadron-side observables}
\label{subsec:HSO}

By integrating over the lepton-side kinematic parameters $E_\pi$, $\theta_\pi$, and $\phi_\pi$, we can obtain the two-fold differential decay rate as follows
\begin{equation}
\frac{d^2\Gamma}{dq^2 d\cos\theta_{\Lambda}} = \frac{1}{2} \frac{d\Gamma}{dq^2} 
\left[1 + \alpha_{\Lambda_c} P_{\Lambda_c}(q^2) \cos\theta_{\Lambda} \right]. 
\end{equation}
Here, $P_{\Lambda_c}(q^2)$ represents the $\Lambda_c$ spin polarization, which is defined as
\begin{equation}
P_{\Lambda_c}(q^2) \equiv \frac{d\Gamma^{\lambda_{\Lambda_c}=1/2}/dq^2 - d\Gamma^{\lambda_{\Lambda_c}=-1/2}/dq^2}{d\Gamma^{\lambda_{\Lambda_c}=1/2}/dq^2 + d\Gamma^{\lambda_{\Lambda_c}=-1/2}/dq^2}.
\end{equation}

The differential decay rates for the polarized intermediate state $\Lambda_c$ baryon are given by
\begin{equation}
\frac{d\Gamma^{\lambda_{\Lambda_c}= \pm 1/2}}{dq^2} = {\cal N}\left(A_0^\pm + A_1^\pm \kappa_\tau + A_2^\pm \kappa_\tau^2 \right), 
\end{equation}
\begin{align*}
A_0^\pm \equiv 
& \frac{3}{2}\left| {\cal A}_{\perp_t} \pm {\cal A}_{\parallel_t} \right|^2 
+ \left| {\cal A}_{\perp_1} \pm {\cal A}_{\parallel_1} \right|^2 
+ \left| {\cal A}_{\perp_0} \pm {\cal A}_{\parallel_0} \right|^2 
\\[2mm]
&+ 2 \left| {\cal A}^T_{\perp_1} \pm {\cal A}^T_{\parallel_1} \right|^2 
+ 2\left| {\cal A}^T_{\perp_0} \pm {\cal A}^T_{\parallel_0}\right|^2,
\\[2mm]
A_1^\pm \equiv
& -6{\rm Re}\left[\left({\cal A}_{\perp_1} \pm {\cal A}_{\parallel_1}\right)\left({\cal A}^T_{\perp_1} \pm {\cal A}^T_{\parallel_1}\right)^* + \left({\cal A}_{\perp_0} \pm {\cal A}_{\parallel_0}\right)\left({\cal A}^T_{\perp_0} \pm {\cal A}^T_{\parallel_0}\right)^*\right],
\\[2mm]
A_2^\pm \equiv
& \frac{1}{2}\left| {\cal A}_{\perp_1} \pm {\cal A}_{\parallel_1} \right|^2
+ \frac{1}{2}\left| {\cal A}_{\perp_0} \pm {\cal A}_{\parallel_0} \right|^2
+ 4\left| {\cal A}^T_{\perp_1} \pm {\cal A}^T_{\parallel_1} \right|^2
+ 4\left| {\cal A}^T_{\perp_0} \pm {\cal A}^T_{\parallel_0}\right|^2,
\end{align*}
where the dimensionless parameter $\kappa_\tau \equiv m_\tau/\sqrt{q^2}$, and the factor
\begin{equation}
{\cal N} \equiv \frac{G_F^2 \left| V_{cb}\right|^2 \left|{\bm p}_{\Lambda_c} \right| q^2}{384 \pi^3 m_{\Lambda_b}^2} \left(1 - \kappa_\tau^2\right)^2 {\cal B}(\tau \to \pi^- \nu_\tau) {\cal B}(\Lambda_c \to \Lambda \pi^+). 
\end{equation}

Further integrating over the variable $\theta_{\Lambda}$, we can obtain the following differential decay rate depending only on $q^2$,
\begin{align}
\frac{d\Gamma}{dq^2} 
&= \frac{d\Gamma^{\lambda_{\Lambda_c}= 1/2}}{dq^2} + \frac{d\Gamma^{\lambda_{\Lambda_c}= - 1/2}}{dq^2} 
\nonumber \\[2mm]
&= {\cal N}\left[\left(A_0^+ + A_0^- \right)  + \left(A_1^+ + A_1^- \right)  \kappa_\tau + \left(A_2^+ + A_2^- \right)  \kappa_\tau^2 \right].
\label{eq:BRq2} 
\end{align}
Our $d\Gamma/dq^2$ (apart from ${\cal B}(\tau \to \pi^- \nu_\tau) {\cal B}(\Lambda_c \to \Lambda \pi^+)$) is consistent with that in Ref.~\cite{Datta:2017aue}, which has also been checked by Refs.~\cite{Hu:2018veh,Boer:2019zmp}. Since we have integrated over all the lepton-side kinematic parameters, the observables constructed above are not affected by $\tau$ decay dynamics, so they are also applicable to light leptons $\ell =\mu,\,e$ (Necessary replacement $m_\tau \to m_\ell$ and removal of factor ${\cal B}(\tau \to \pi^- \nu_\tau)$ are required). The universality of lepton flavor can be tested by comparing the predicted values of observables $d\Gamma/dq^2$ or $P_{\Lambda_c}(q^2)$ of $\tau$ and $\ell$.

\subsection{Lepton-side observables}
\label{subsec:LSO}

By integrating over the hadron-side kinematic parameters $\theta_{\Lambda}$, and one or two lepton-side kinematic parameters, we can construct a variety of observables. These observables depend on at least one kinematic parameter of $\pi^-$, so they only exist in $\tau$ channels, and specifically for the $\tau \to \pi^- \nu_\tau$ decay.

The differential decay rates for which $E_\pi$ has not been integrated over can be expressed simply as the angular observables ${\cal K}_i$. To reduce the uncertainty of theoretical predictions, we use $d^2\Gamma/(dq^2 dE_\pi)$ to normalize them.
\begin{align}
\frac{d^3\Gamma}{dq^2 dE_\pi d\cos\theta_\pi} &= \frac{3}{2} \frac{d^2\Gamma}{dq^2 dE_\pi} \frac{{\cal K}_{1ss} \sin^2\theta_\pi + {\cal K}_{1cc} \cos^2\theta_\pi + {\cal K}_{1c} \cos\theta_\pi}{2 {\cal K}_{1ss} + {\cal K}_{1cc}},
\label{eq:d3thepi}
\\[2mm]
\frac{d^3\Gamma}{dq^2 dE_\pi d\phi_\pi} &= \frac{1}{2\pi} \frac{d^2\Gamma}{dq^2 dE_\pi} \left(1+ \frac{3\pi^2}{16} \frac{{\cal K}_{3s} \sin\phi_\pi + {\cal K}_{4s} \cos\phi_\pi}{2 {\cal K}_{1ss} + {\cal K}_{1cc}} \right), 
\label{eq:d3phipi}
\end{align}
with
\begin{equation}
\frac{d^2\Gamma}{dq^2 dE_\pi} = \frac{d\Gamma}{dq^2} \frac{4 \kappa_\tau^2}{\sqrt{q^2} \left(\kappa_\tau^2 - \kappa_\pi^2 \right)^2 \left(1 - \kappa_\tau^2\right)^2} \frac{2 {\cal K}_{1ss} + {\cal K}_{1cc}}{\left(A_0^+ + A_0^- \right)  + \left(A_1^+ + A_1^- \right)  \kappa_\tau + \left(A_2^+ + A_2^- \right)  \kappa_\tau^2}.
\end{equation}
The forward-backward asymmetry of the $\pi^-$ meson can be obtained by the difference between the integrals of the Eq.~\eqref{eq:d3thepi} on the interval $[0,\pi/2)$ and $[\pi/2,\pi)$. We can define the following asymmetry $A_{FB} (q^2,E_\pi)$ as a function of $q^2$ and $E_\pi$,
\begin{align}
A_{FB} (q^2,E_\pi) =& \frac{\int_{0}^{1} \frac{d^3\Gamma}{dq^2 dE_\pi d\cos\theta_\pi} d\cos\theta_\pi - \int_{-1}^{0} \frac{d^3\Gamma}{dq^2 dE_\pi d\cos\theta_\pi} d\cos\theta_\pi}{\frac{d^2\Gamma}{dq^2 dE_\pi}}
\nonumber\\[2mm]
=& \frac{3}{2} \frac{{\cal K}_{1c}}{2 {\cal K}_{1ss} + {\cal K}_{1cc}}.
\label{eq:AFBq2Epi}
\end{align}
The difference between the integrals of the Eq.~\eqref{eq:d3phipi} on the interval $[0,\pi)$ and $[\pi,2\pi)$ can isolate the angular observable ${\cal K}_{3s}$, which is nonzero only if the NP induces a complex contribution to the amplitude.

Further integrating over the $\pi^-$ energy $E_\pi$ in Eqs.~\eqref{eq:d3thepi}
and \eqref{eq:d3phipi}, one can obtain the two-fold differential decay rates $d^2\Gamma/(dq^2d\cos\theta_\pi)$ and $d^2\Gamma/(dq^2d\phi_\pi)$. Similarly, we can use them to construct asymmetry observables that do not depend on the variable $E_\pi$. For example, the forward-backward asymmetry of the $\pi^-$ meson as a function of $q^2$ can be defined as
\begin{equation}
A_{FB} (q^2) = \frac{\int_{0}^{1} \frac{d^2\Gamma}{dq^2 d\cos\theta_\pi} d\cos\theta_\pi - \int_{-1}^{0} \frac{d^2\Gamma}{dq^2 d\cos\theta_\pi} d\cos\theta_\pi}{\frac{d\Gamma}{dq^2}}.
\label{eq:AFB}
\end{equation}
This result can also be obtained by integrating over $E_\pi$ separately in the numerator and denominator in Eq.~\eqref{eq:AFBq2Epi}.

\subsection{The angular observables $\widehat{{\cal K}}_i (q^2)$ and $\widehat{{\cal K}}_i$}
\label{subsec:tenAO}

Starting from five-fold differential decay rate~\eqref{eq:5fold}, integrating over the variable $E_\pi$, and after proper normalization, we can obtain the following angular function
\begin{align}
\widehat{{\cal K}}\left( q^2,\cos\theta_{\Lambda}, \cos\theta_\pi, \phi_\pi\right)  
\equiv
&\frac{\int \frac{d^5\Gamma}{dq^2 dE_\pi d\cos\theta_\pi d\phi_\pi d\cos\theta_\Lambda} dE_\pi}{\int \frac{d^2\Gamma}{dq^2 dE_\pi} dE_\pi}
\nonumber \\[2mm]
= &\frac{3}{8\pi}\sum_{i=1}^{10} \widehat{{\cal K}}_i(q^2) \Omega_i(\cos\theta_{\Lambda}, \cos\theta_\pi, \phi_\pi),
\end{align}
with the angular observables $\widehat{{\cal K}}_i (q^2)$ given by
\begin{equation}
\widehat{{\cal K}}_i(q^2) \equiv \frac{\int {\cal K}_i(q^2, E_\pi) dE_\pi}{\int \left( 2{\cal K}_{1ss} + {\cal K}_{1cc} \right) dE_\pi}.
\label{eq:Kiq2}
\end{equation}
Comparing Eqs.~\eqref{eq:AFBq2Epi}, \eqref{eq:AFB} and \eqref{eq:Kiq2}, we can easily get the $A_{FB}(q^2) = \frac{3}{2} \widehat{{\cal K}}_{1c}(q^2)$. The spin polarization of $\Lambda_c$ baryon satisfies $\alpha_{\Lambda_c} P_{\Lambda_c}(q^2) = 2 \widehat{{\cal K}}_{2ss}(q^2) + \widehat{{\cal K}}_{2cc}(q^2)$.

If the variables $E_\pi$ and $q^2$ in Eq.~\eqref{eq:5fold} are simultaneously integrated over, we can obtain the following angular distribution
\begin{align}
\widehat{{\cal K}}\left(\cos\theta_{\Lambda}, \cos\theta_\pi, \phi_\pi\right)  
\equiv
&\frac{\int \frac{d^5\Gamma}{dq^2 dE_\pi d\cos\theta_\pi d\phi_\pi d\cos\theta_\Lambda} dE_\pi dq^2}{\int \frac{d^2\Gamma}{dq^2 dE_\pi} dE_\pi dq^2}
\nonumber \\[2mm]
= &\frac{3}{8\pi}\sum_{i=1}^{10} \widehat{{\cal K}}_i \Omega_i(\cos\theta_{\Lambda}, \cos\theta_\pi, \phi_\pi),
\end{align}
with the angular observables $\widehat{{\cal K}}_i$ given by
\begin{equation}
\widehat{{\cal K}}_i \equiv \frac{\int (q^2)^{3/2}\sqrt{Q_+ Q_-}{\cal K}_i(q^2, E_\pi) dE_\pi dq^2}{\int (q^2)^{3/2}\sqrt{Q_+ Q_-}\left( 2{\cal K}_{1ss} + {\cal K}_{1cc} \right) dE_\pi dq^2}.
\label{eq:Ki}
\end{equation}
Our choice of the normalization in Eq.~\eqref{eq:Kiq2} and \eqref{eq:Ki} make the first two angular observables exactly satisfy the relationships, $2\widehat{{\cal K}}_{1ss}(q^2) + \widehat{{\cal K}}_{1cc}(q^2) = 1$ and $2\widehat{{\cal K}}_{1ss} + \widehat{{\cal K}}_{1cc} = 1$. All observables related to $\Lambda_b^0 \to \Lambda_c^+ (\to \Lambda^0 \pi^+)\tau^- (\to \pi^- \nu_\tau)\bar{\nu}_\tau$ decay can be expressed linearly by the corresponding angular observables ${\cal K}_i$ and the normalized ones $\widehat{{\cal K}}_i$, which contain $0\sim 2$ variables in $E_\pi$ and $q^2$. For instance, the forward-backward asymmetry of the $\pi^-$ meson is $A_{FB} = \frac{3}{2} \widehat{{\cal K}}_{1c}$, or as a function of $q^2$ is $A_{FB} (q^2) = \frac{3}{2} \widehat{{\cal K}}_{1c}(q^2)$, or as a function of $q^2$ and $E_\pi$ is $A_{FB} (q^2,E_\pi) = \frac{3}{2} \widehat{{\cal K}}_{1c}(q^2,E_\pi)$. In the following numerical analyses, we only focus on the normalized angular observables $\widehat{{\cal K}}_i$ and $\widehat{{\cal K}}_i(q^2)$, because they have less theoretical uncertainty to facilitate the discussion of the effects of the NP. The $\widehat{{\cal K}}_i(q^2, E_\pi)$ are not included for the time being, as they require more experimental statistics than $\widehat{{\cal K}}_i$ and $\widehat{{\cal K}}_i(q^2)$. When the statistics are large enough, it is necessary to discuss $\widehat{{\cal K}}_i(q^2, E_\pi)$ in detail.

%%%%%%%%%%%%%%%%%%%%%%%%%%%%%%%%%
\begin{figure}[h]
\centering
\includegraphics[width=0.35\textwidth]{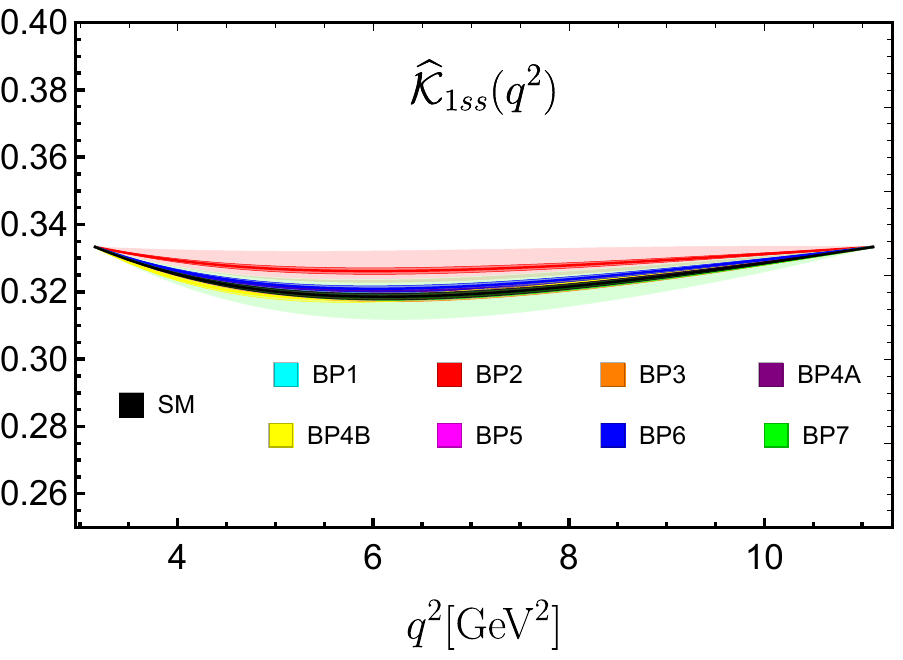}
\\
\includegraphics[width=0.3\textwidth]{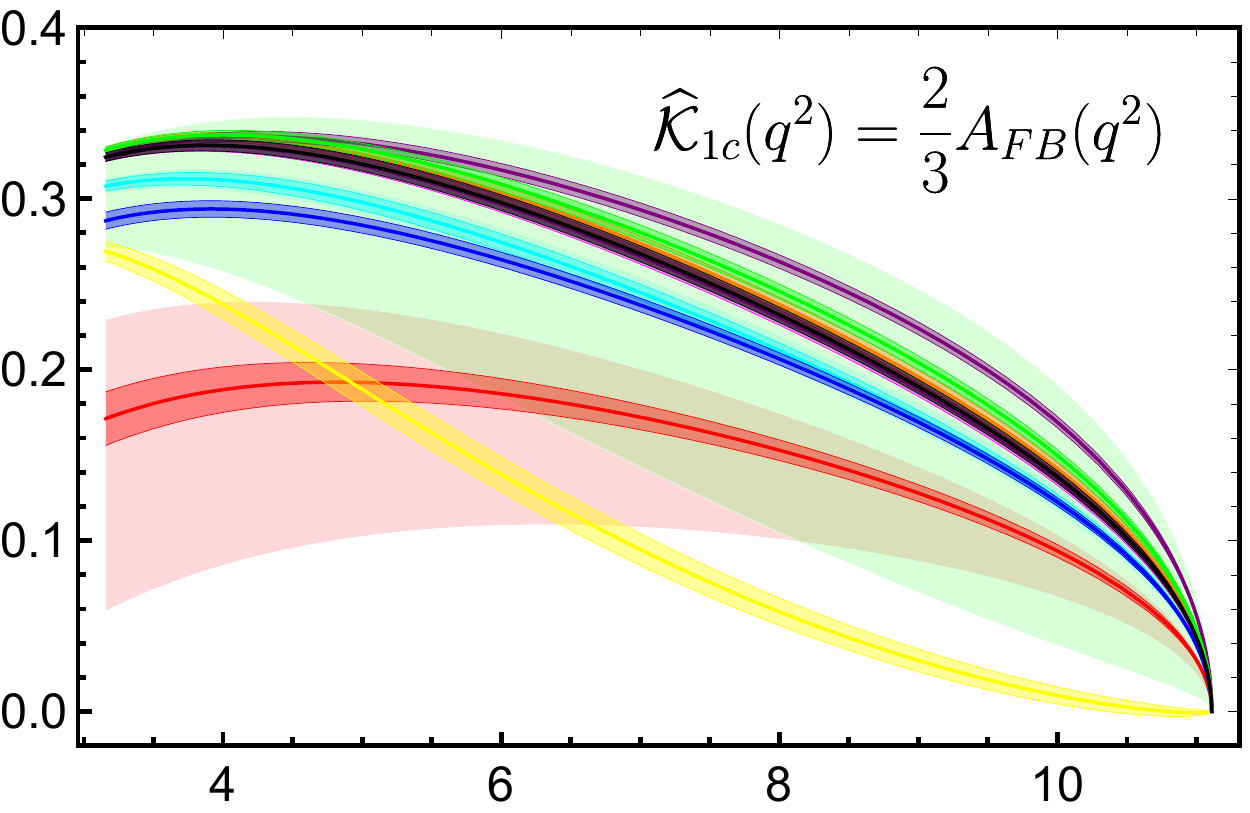}
\quad
\includegraphics[width=0.3\textwidth]{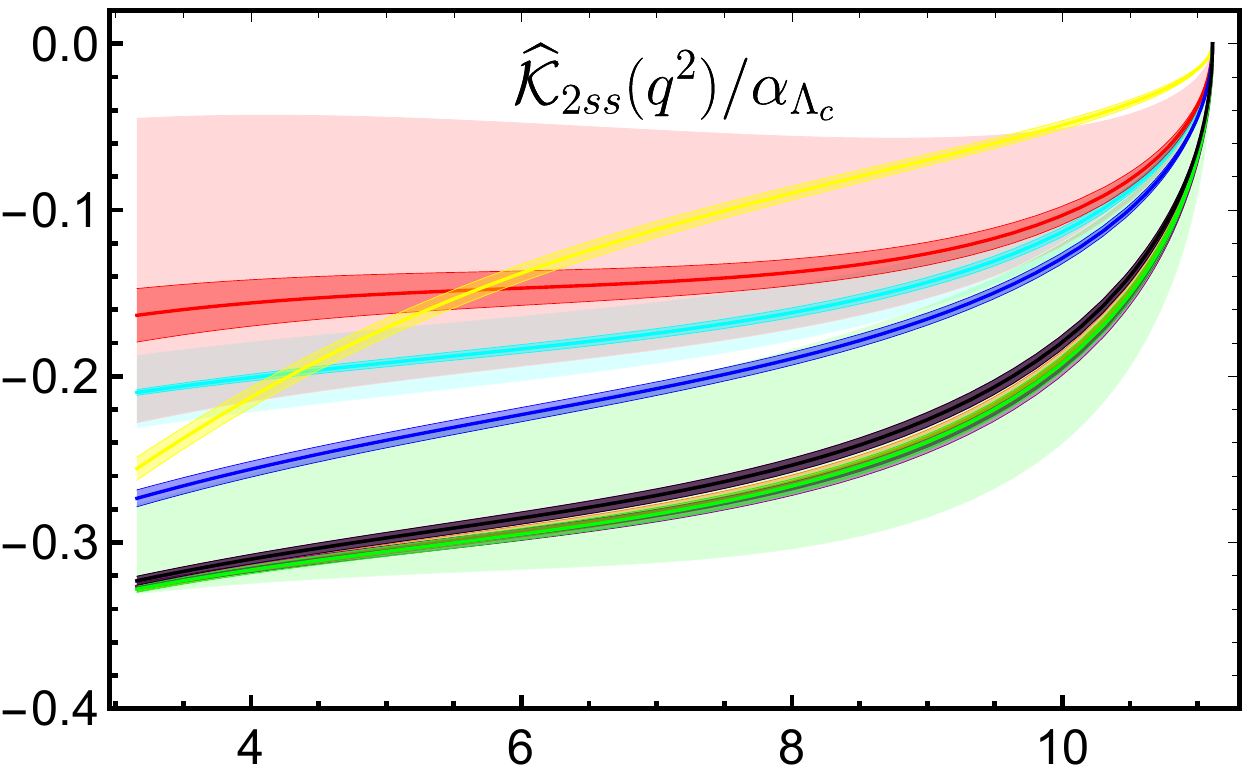}
\\
\includegraphics[width=0.3\textwidth]{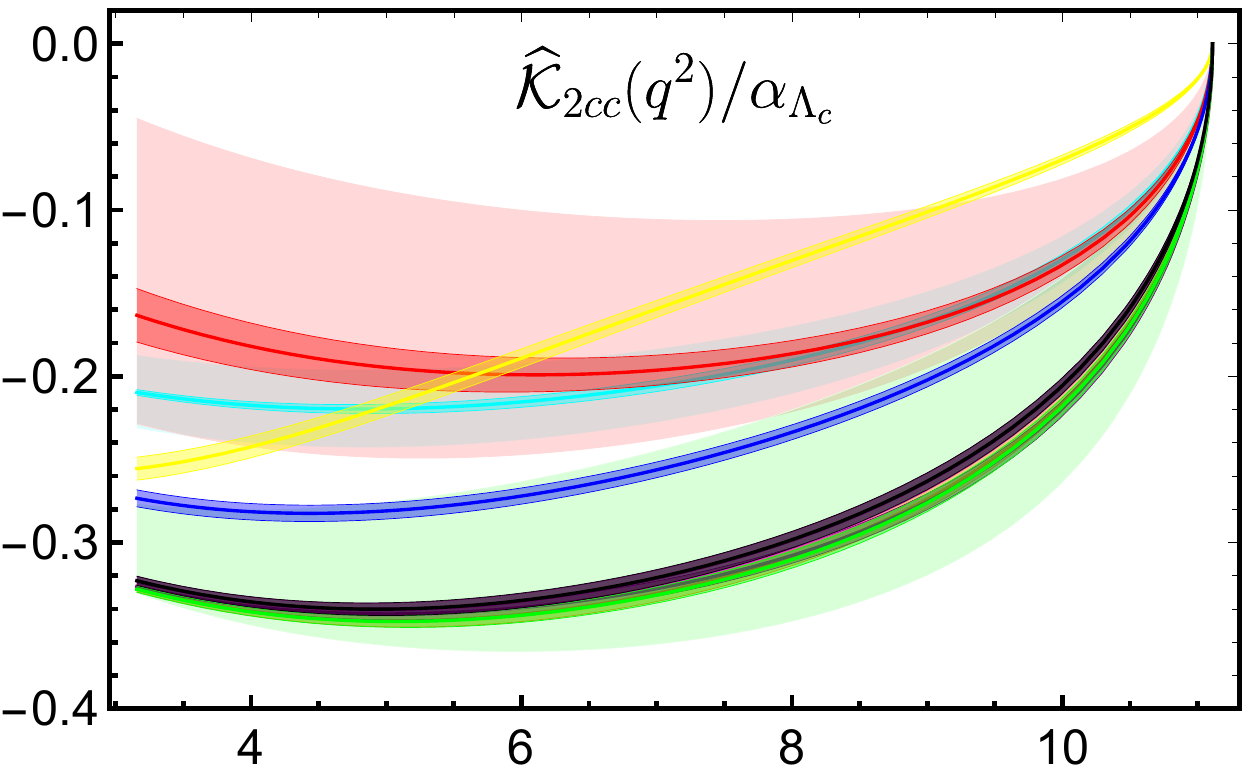}
\quad
\includegraphics[width=0.3\textwidth]{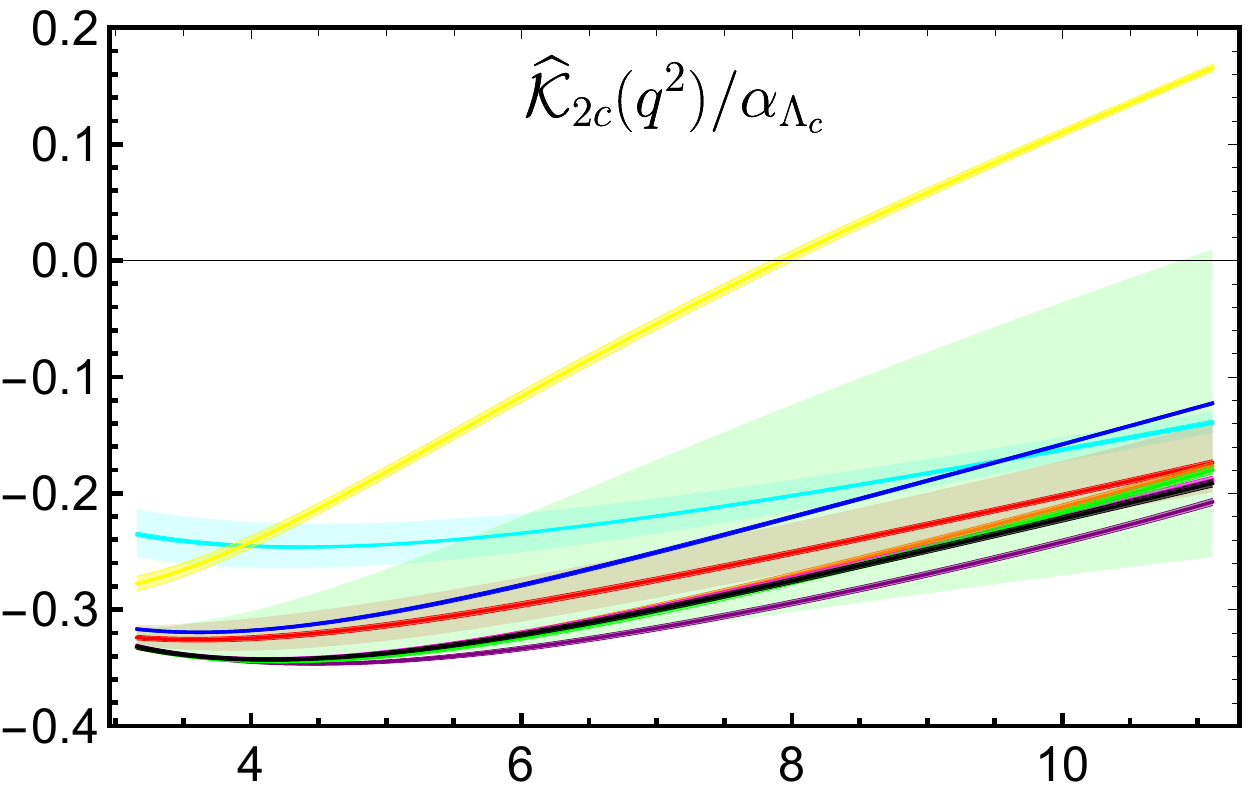}
\\
\includegraphics[width=0.3\textwidth]{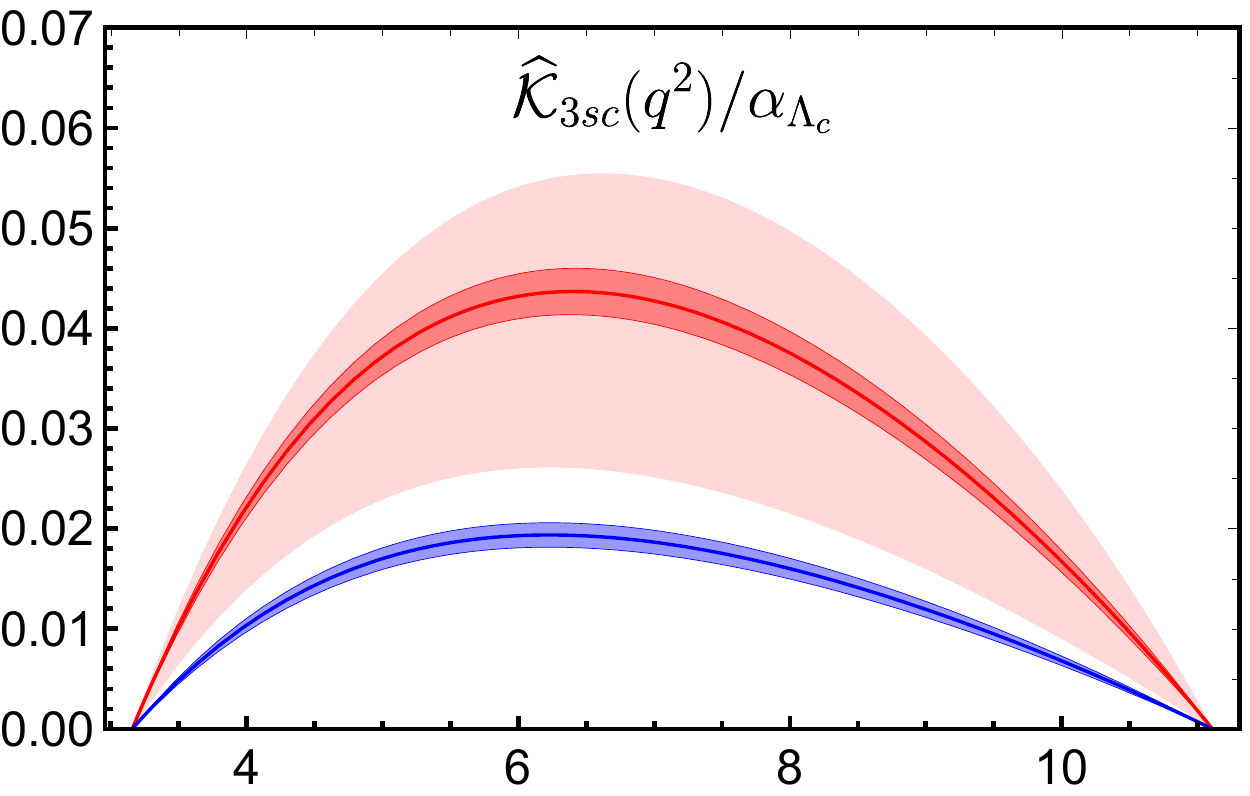}
\quad
\includegraphics[width=0.3\textwidth]{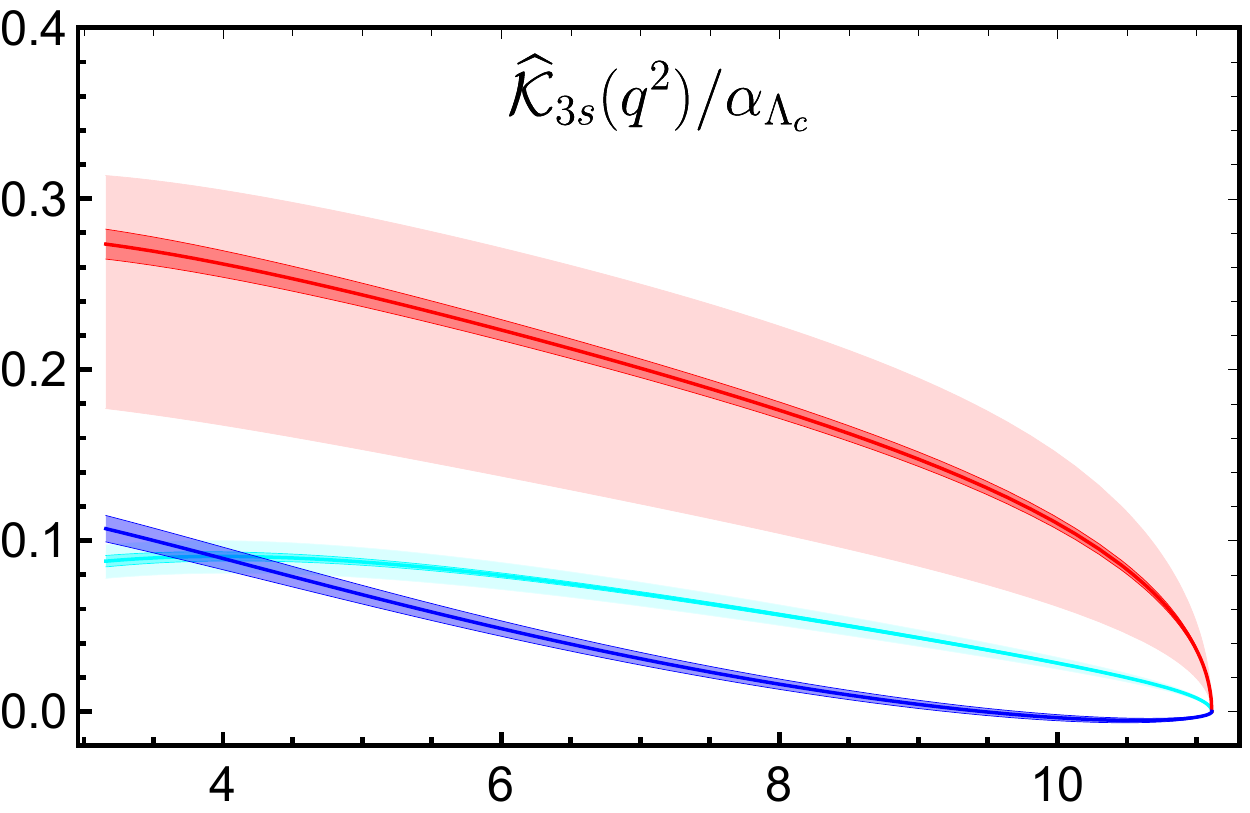}
\\
\includegraphics[width=0.3\textwidth]{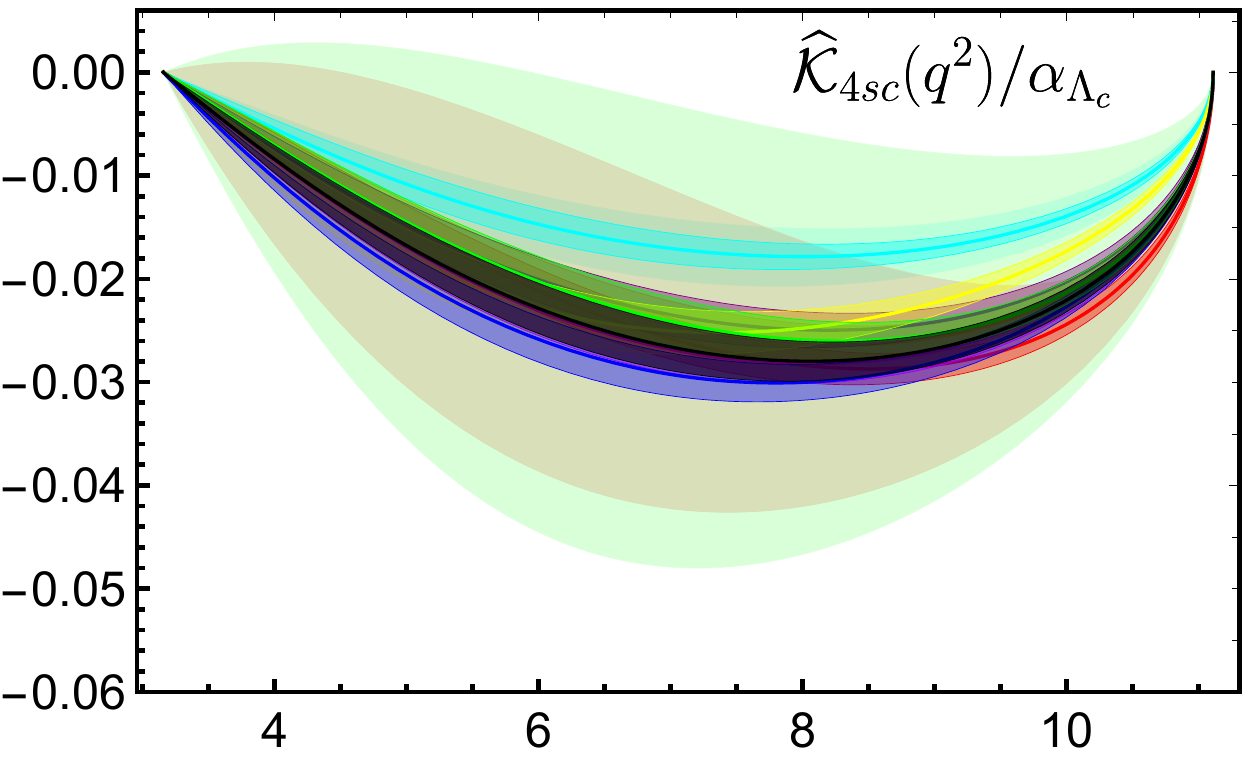}
\quad
\includegraphics[width=0.3\textwidth]{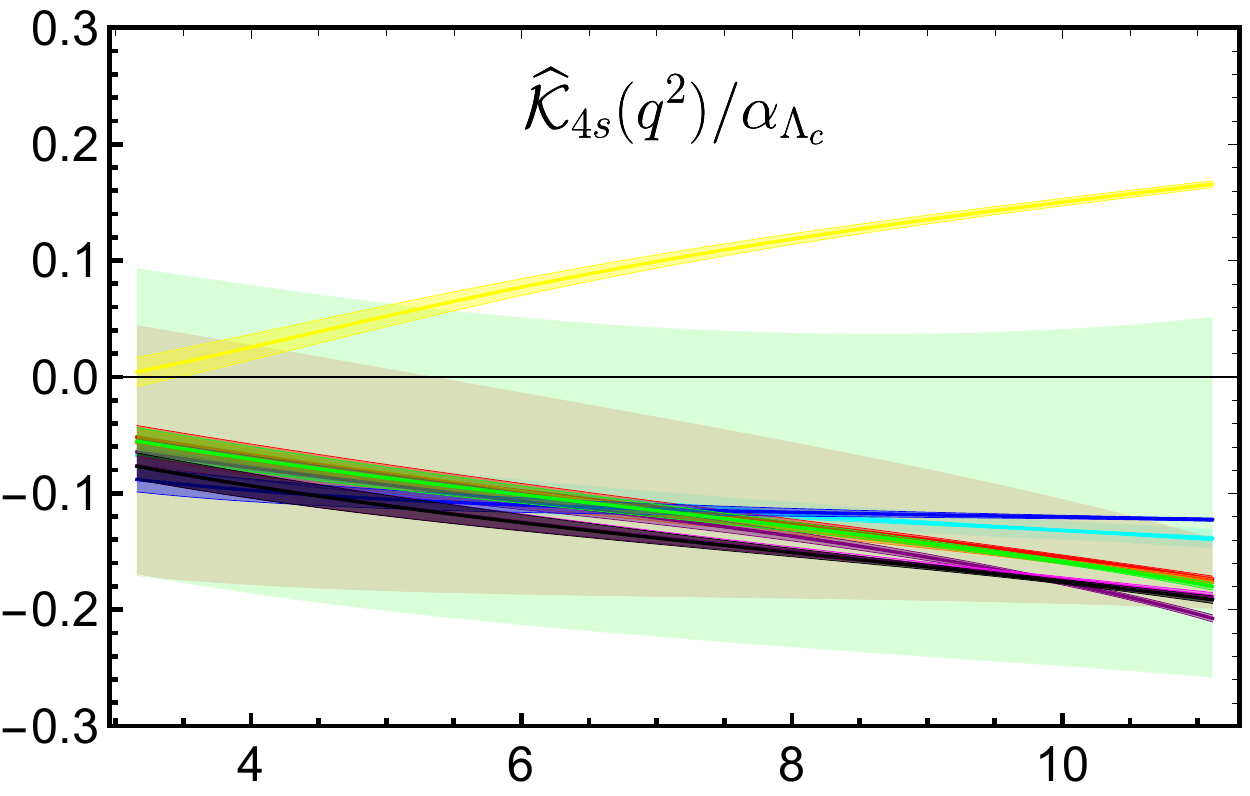}
\caption{\label{fig:angularObs} \small The angular observables $\widehat{\cal K}_i (q^2)$ as a function of $q^2$, predicted both within the SM and in some NP scenarios. The asymmetry parameter $\alpha_{\Lambda_c}$ is factored out in $\widehat{\cal K}_i (q^2)$ with $i=2ss,\, 2cc,\, 2c,\, 3sc,\, 3s,\, 4sc$, and $4s$. The observable $\widehat{{\cal K}}_{1cc}(q^2)$ can be obtained as $1-2\widehat{{\cal K}}_{1ss}(q^2)$. The width of each curve is derived from the theoretical uncertainties of $\Lambda_b \to \Lambda_c$ form factors, except that the widths of light-colored curves are derived from the uncertainties of both the form factors and the NP parameters given in BP1, BP2, and BP7.}
\end{figure}
%%%%%%%%%%%%%%%%%%%%%%%%%%%%%%%%%

\section{Numerical results}
\label{sec:numerical}

The model-independent analyses to study the NP effects in $\bar{B}\to D^{(*)} \tau^- \bar{\nu}_\tau$ decays have been completed in the previous literatures~\cite{Alok:2017qsi,Hu:2018veh,Alok:2019uqc,Murgui:2019czp,Blanke:2018yud,Blanke:2019qrx,Cheung:2020sbq,Kumbhakar:2020jdz}. In order to illustrate our results numerically, we select a variety of best-fit values as the NP scenarios. These best-fit values are usually performed in a set of bases, which is equivalent to Eq.~\eqref{eq:Heff} by the following relations
\begin{align}
	g_V =& 1+C_{V_L}+C_{V_R}, & g_A =& -1-C_{V_L}+C_{V_R}, & \nonumber\\[2mm]
	g_S =& C_{S_L} + C_{S_R}, & g_P =& -C_{S_L} + C_{S_R}, & g_T = C_T.
\end{align}
We should choose the best-fit values from recent global analyses~\cite{Alok:2019uqc,Murgui:2019czp,Blanke:2019qrx,Cheung:2020sbq,Kumbhakar:2020jdz}, including two new results recently announced by the Belle experiment: the first measurement~\cite{Abdesselam:2019wbt} of $D^*$ longitudinal polarization fraction in $B \to D^* \tau \nu$ decay and the new measurements~\cite{Belle:2019rba} of $R(D^{(*)})$. The fitting results show that a single Wilson coefficient $C_{V_L}$ can explain the experimental data well. However, in this scenario, there is no change in the normalized angular observables $\widehat{{\cal K}}_i$ and $\widehat{{\cal K}}_i(q^2)$, so we do not choose it.  The scenario with a single $C_{V_R}$ is allowed only if $C_{V_R}$ is complex. We choose $\left({\rm Re}\left[C_{V_R}\right], \, {\rm Im}\left[C_{V_R}\right]\right) = \left(-0.031(34),\,  0.460(52)\right)$  (with correlation 0.59)~\cite{Cheung:2020sbq} as the NP scenario and mark it as BP1\footnote{The corresponding complex conjugate fitting value $\left({\rm Re}\left[C_{V_R}\right], \, {\rm Im}\left[C_{V_R}\right]\right) = \left(-0.031(34),\,  -0.460(52)\right)$ (with correlation $-0.59$)~\cite{Cheung:2020sbq} is marked as BP1$^*$. Similar conventions are used in other two complex scenarios BP2$^*$ and BP6$^*$.}. The scenario with a single $C_{S_L}$ or $C_{S_R}$ is ruled out by the branching ratio of $B_c \to \tau \nu$ decay~\cite{Li:2016vvp,Celis:2016azn,Alonso:2016oyd}. For the scenario with a single $C_{T}$, we take $\left({\rm Re}\left[C_{T}\right], \, {\rm Im}\left[C_{T}\right]\right) = \left(0.011(62),\,  0.164(60)\right)$ (with correlation 0.98)~\cite{Cheung:2020sbq} as the other NP scenario and mark it as BP2. 

In Ref.~\cite{Blanke:2019qrx}, a set of benchmark points is determined by considering the best-fit points of different scenarios induced by specific UV models. Specifically, we will choose the best-fit points of the following four different NP hypotheses, all of which can explain the $R(D^{(*)})$ anomalies, as our NP benchmark points (the remaining Wilson coefficients $C_i$ are set to zero in each case)
\begin{align*}
\text{BP3:} \qquad &  \left(C_{V_L},\, C_{S_L} = -4 C_T\right)= \left(0.10, \, -0.04\right)
\\[2mm]
\text{BP4:} \qquad  &  \left(C_{S_R},\, C_{S_L}\right)= \left(0.21, \, -0.15\right)\, \text{(A)} \quad \text{or}  \quad  \left(-0.26, \, -0.61\right)\, \text{(B)}
\\[2mm]
\text{BP5:} \qquad &  \left(C_{V_L},\, C_{S_R} \right)= \left(0.08, \, -0.01\right)
\\[2mm]
\text{BP6:} \qquad &  \left({\rm Re}\left[C_{S_L} = 4 C_T\right],\, {\rm Im}\left[C_{S_L} = 4 C_T\right] \right)= \left(-0.06, \, 0.31\right)
\end{align*}
where the Wilson coefficients are given at the scale $\mu = 1$~TeV, and we run them down to the typical energy scale $\mu = m_b$~\cite{Hu:2018veh,Blanke:2018yud,Gonzalez-Alonso:2017iyc}. Finally, we choose a set of values labelled ``Min 1b" in Table 8 of Ref.~\cite{Murgui:2019czp} as our BP7 scenario
\begin{equation*}
\text{BP7:} \qquad   \left(C_{V_L},\, C_{S_R}, \, C_{S_L}, \, C_T \right)= \left(0.09^{+0.13}_{-0.12}, \, 0.086^{+0.12}_{-0.61},\, -0.14^{+0.52}_{-0.07},\, 0.008^{+0.046}_{-0.044}\right)
\end{equation*}

Next, we should discuss the entire set of angular observables, including the functions $\widehat{\cal K}_i (q^2)$ and the numbers $\widehat{\cal K}_i$, within the SM and in these NP scenarios respectively. For angular observables ${\cal K}_i$ with $i=2ss,\, 2cc,\, 2c,\, 3sc,\, 3s,\, 4sc$, and $4s$, we factor out the asymmetry parameter $\alpha_{\Lambda_c} = -0.82 \pm 0.09$~\cite{Boer:2019zmp}, because it can bring great uncertainty to these observables, thus interfering with the emergence of the NP effects.

%%%%%%%%%%%%%%%%%%%%%%%%%%%%%%%%%%%%%%%%%%%%
\begin{table}
\begin{center}
\begin{tabular}{cccccc} 
\hline
observable & SM & BP1 & BP2 & BP3 & BP4A  
\\ \hline
$\widehat{\cal K}_{1ss}$ & 0.323(1) & 0.323(1)(0) & $0.328(1)^{+0.004}_{-0.002}$ & 0.323(1) & 0.324(1)
\\ 
$\widehat{\cal K}_{1c}$ & 0.224(5) & 0.204(4)(4) & $0.145(6)^{+0.016}_{-0.043}$ & 0.230(4) & 0.251(4)
\\
$\widehat{\cal K}_{2ss}/\alpha_{\Lambda_c}$ & $-0.240(4)$ & $-0.153(3)(14)$ & $-0.130(8)^{+0.067}_{-0.027}$ & $-0.249(4)$ & $-0.254(4)$
\\
$\widehat{\cal K}_{2cc}/\alpha_{\Lambda_c}$ & $-0.279(5)$ & $-0.178(3)(16)$ & $-0.172(8)^{+0.066}_{-0.027}$ & $-0.289(5)$ & $-0.289(5)$
\\
$\widehat{\cal K}_{2c}/\alpha_{\Lambda_c}$ & $-0.274(3)$ & $-0.200(1)(13)$ & $-0.252(3)^{+0.020}_{-0.013}$ & $-0.270(3)$ & $-0.291(3)$
\\
$\widehat{\cal K}_{3sc}/\alpha_{\Lambda_c}$ & 0 & 0 & $0.032(2)^{+0.008}_{-0.012}$ & 0 & 0
\\
$\widehat{\cal K}_{3s}/\alpha_{\Lambda_c}$ & 0 & 0.055(1)(5) & $0.172(5)^{+0.045}_{-0.067}$ & 0 & 0
\\
$\widehat{\cal K}_{4sc}/\alpha_{\Lambda_c}$ & $-0.024(2)$ & $-0.015(1)(1)$ & $-0.024(1)^{+0.009}_{-0.010}$ & $-0.022(2)$ & $-0.021(1)$
\\
$\widehat{\cal K}_{4s}/\alpha_{\Lambda_c}$ & $-0.149(4)$ & $-0.117(3)(8)$ & $-0.124(4)^{+0.065}_{-0.063}$ & $-0.130(4)$ & $-0.138(4)$ 
\\ 
\hline \hline
observable & BP4B & BP5 & BP6 & BP7 &
\\ \hline
$\widehat{\cal K}_{1ss}$  & 0.324(1) & 0.323(1) & 0.325(1) & $0.323(1)^{+0.004}_{-0.004}$ &
\\ 
$\widehat{\cal K}_{1c}$ & 0.067(7) & 0.222(5) & 0.198(4) & $0.237(4)^{+0.035}_{-0.125}$ & 
\\
$\widehat{\cal K}_{2ss}/\alpha_{\Lambda_c}$ & $-0.091(4)$ & $-0.240(4)$ & $-0.181(4)$ & $-0.252(4)^{+0.075}_{-0.034}$ & 
\\
$\widehat{\cal K}_{2cc}/\alpha_{\Lambda_c}$ & $-0.126(4)$ & $-0.279(5)$ & $-0.219(4)$ & $-0.291(5)^{+0.082}_{-0.032}$ & 
\\
$\widehat{\cal K}_{2c}/\alpha_{\Lambda_c}$ & 0.004(4) & $-0.272(3)$ & $-0.220(2)$ & $-0.274(3)^{+0.146}_{-0.025}$ & 
\\
$\widehat{\cal K}_{3sc}/\alpha_{\Lambda_c}$ & 0 & 0 & 0.014(1) & 0 & 
\\
$\widehat{\cal K}_{3s}/\alpha_{\Lambda_c}$ & 0 & 0 & 0.022(3) & 0 & 
\\
$\widehat{\cal K}_{4sc}/\alpha_{\Lambda_c}$ & $-0.021(2)$ & $-0.024(2)$ & $-0.025(1)$ & $-0.022(2)^{+0.016}_{-0.016}$ & 
\\
$\widehat{\cal K}_{4s}/\alpha_{\Lambda_c}$ & 0.117(5) & $-0.148(4)$ & $-0.115(4)$ & $-0.128(4)^{+0.166}_{-0.100}$ & 
\\ 
\hline
\end{tabular}
\caption{\label{tab:numres} \small Predictions for the entire set of angular observables $\widehat{\cal K}_i$ within the SM and in some NP scenarios. The asymmetry parameter $\alpha_{\Lambda_c}$ is factored out in $\widehat{\cal K}_i$ with $i=2ss,\, 2cc,\, 2c,\, 3sc,\, 3s,\, 4sc$, and $4s$. The observable $\widehat{{\cal K}}_{1cc}$ can be obtained as $1-2\widehat{{\cal K}}_{1ss}$. The first uncertainties come from the $\Lambda_b \to \Lambda_c$ form factors, and the second (only in BP1, BP2, and BP7) come from the NP parameters. }
\end{center}
\end{table}
%%%%%%%%%%%%%%%%%%%%%%%%%%%%%%%%%%%%%%%%%%%%

In our numerical analyses, we use the $\Lambda_b \to \Lambda_c$ form factors computed in lattice QCD including all the types of Lorentz structures of the NP effective operators~\cite{Detmold:2015aaa,Datta:2017aue}. The results of the angular observables $\widehat{\cal K}_i (q^2)$ as a function of $q^2$ are shown in Figure~\ref{fig:angularObs}. When only the central value of Wilson coefficients in each NP scenario is considered (corresponding to the normally colored regions in Figure~\ref{fig:angularObs}), the theoretical uncertainties of the observables $\widehat{\cal K}_i (q^2)$ mainly come from the $\Lambda_b \to \Lambda_c$ form factors since the cancellations through normalization to the decay rate and the asymmetry parameter $\alpha_{\Lambda_c}$ has been factored out. Benefiting from the correlation between the uncertainties of the $\Lambda_b \to \Lambda_c$ form factors, these observables have small uncertainties. The accurate prediction of the observables corresponding to each NP point enables us to use them to discuss the NP effects. In Figure~\ref{fig:angularObs}, we also give the predictions of the observables $\widehat{\cal K}_i (q^2)$ as the NP Wilson coefficients in BP1, BP2, and BP7 vary within $1\sigma$ level (corresponding to the light-colored regions which simultaneously contain uncertainties of the form factors and the NP Wilson coefficients). Let us now comment on the results we obtain.
\begin{itemize}
\item \underline{$\widehat{\cal K}_{1ss} (q^2)$}:  At two endpoints, the values of $\widehat{\cal K}_{1ss} (q^2)$ are fixed at $\frac{1}{3}$.  Specifically, at endpoint $q^2_{\rm max} = (m_{\Lambda_b} - m_{\Lambda_c})^2$, the transversity amplitudes ${\cal A}_{\parallel_t}$, ${\cal A}_{\perp_{1,0}}$, and ${\cal A}^T_{\perp_{1,0}}$ are vanishing. The endpoint relations of the helicity form factors $G_+(q^2_{\rm max}) = G_\perp(q^2_{\rm max})$~\cite{Detmold:2015aaa} and $\widetilde{h}_+(q^2_{\rm max}) = \widetilde{h}_\perp(q^2_{\rm max})$~\cite{Datta:2017aue} result in $\widehat{\cal K}_{1ss} (q^2_{\rm max}) = \frac{1}{3}$. Near the endpoint $q^2_{\rm min} = m^2_\tau$, the dimensionless factors after integrating over $E_\pi$ satisfy the following asymptotic relationships
\begin{align}
\bar{S}_t =& \eta \varepsilon^2 + {\cal O}(\varepsilon^3),
&
\bar{S}_1 =& \eta \varepsilon^2 + {\cal O}(\varepsilon^3),
&
\bar{S}_2 =& -\eta \varepsilon^2 + {\cal O}(\varepsilon^3),
\\[2mm]
\bar{S}_3 =&  {\cal O}(\varepsilon^3),
&
\bar{S}^T_1 =& 4\eta \varepsilon^2 + {\cal O}(\varepsilon^3),
&
\bar{S}^T_2 =& -4\eta \varepsilon^2 + {\cal O}(\varepsilon^3),
\\[2mm]
\bar{S}^T_3 =&  {\cal O}(\varepsilon^3),
&
\bar{R}_t =& -\sqrt{2} \eta \varepsilon^2 + {\cal O}(\varepsilon^3),
&
\bar{R}^T_t =& 2\sqrt{2} \eta \varepsilon^2 + {\cal O}(\varepsilon^3),
\\[2mm]
\bar{R}_1 =& -4 \eta \varepsilon^2 + {\cal O}(\varepsilon^3),
&
\bar{R}_2 =& 4 \eta \varepsilon^2 + {\cal O}(\varepsilon^3),
&
\bar{R}_3 =&  {\cal O}(\varepsilon^3),
\end{align}
where $\bar{S}_i (\bar{R}_i) \equiv \int S_i (R_i) dE_\pi$, $\eta \equiv \sqrt{q^2}(1-\kappa_\pi^2)^2$, and $\varepsilon \equiv 1- \kappa_\tau$. By comparing Eq.~\eqref{eq:K1ss} with Eq.~\eqref{eq:K1cc}, one can immediately get $\widehat{\cal K}_{1ss} (q^2_{\rm min}) = \frac{1}{3}$. The NP does not have much impact on $\widehat{\cal K}_{1ss} (q^2)$, even though the uncertainties of the NP parameters are taken into account in scenarios BP1, BP2, and BP7. The NP effect in scenario BP2 contributes the most to $\widehat{\cal K}_{1ss} (q^2)$, but it can only increase $\widehat{\cal K}_{1ss} (q^2)$ by about 3\%.

\item \underline{$\widehat{\cal K}_{1c} (q^2) = \frac{2}{3} A_{FB}(q^2)$}: This observable can clearly distinguish the two best-fit points in the NP hypothesis $\left(C_{S_R},\, C_{S_L}\right)$, which is motivated by models with extra charged Higgs. The predicted value of $\widehat{\cal K}_{1c} (q^2)$ decreases greatly in BP4B, but increases slightly in BP4A. Although there is a great deal of uncertainty, the NP effect in scenario BP2 can still greatly reduce $\widehat{\cal K}_{1c} (q^2)$. The uncertainties of the NP parameters do not bring considerable uncertainty to the prediction of $\widehat{\cal K}_{1c} (q^2)$ in scenario BP1. The predicted values of $\widehat{\cal K}_{1c} (q^2)$ in scenarios BP1 and BP2 do not overlap. In scenarios BP1 and BP6, the predicted values of $\widehat{\cal K}_{1c} (q^2)$ decrease slightly. At the endpoint $q^2_{\rm max}$, the disappearance of  transversity amplitudes ${\cal A}_{\parallel_t}$, ${\cal A}_{\perp_{1,0}}$, and ${\cal A}^T_{\perp_{1,0}}$ leads to ${\cal K}_{1c}(q^2, E_\pi)=0$ (see Eq.~\eqref{eq:K1c}) and thus $\widehat{\cal K}_{1c} (q^2_{\rm max})=0$.

\item \underline{$\widehat{\cal K}_{2ss}(q^2)/\alpha_{\Lambda_c}$}: This observable can also clearly distinguish the two best-fit points in the NP hypothesis $\left(C_{S_R},\, C_{S_L}\right)$. The predicted value of $\widehat{\cal K}_{2ss}(q^2)/\alpha_{\Lambda_c}$ increases greatly in BP4B, but decreases slightly in BP4A. The NP effects in scenarios BP1, BP2, and BP6 can significantly increase the predicted value of this observable. The corresponding region of scenario BP2 almost cover the region of scenario BP1, but these two regions do not coincide with that of scenario BP7. At the endpoint $q^2_{\rm max}$, the disappearance of  transversity amplitudes ${\cal A}_{\parallel_t}$, ${\cal A}_{\perp_{1,0}}$, and ${\cal A}^T_{\perp_{1,0}}$ leads to ${\cal K}_{2ss}(q^2, E_\pi)=0$ (see Eq.~\eqref{eq:K2ss}) and thus $\widehat{\cal K}_{2ss}(q^2_{\rm max})=0$.

\item \underline{$\widehat{\cal K}_{2cc}(q^2)/\alpha_{\Lambda_c}$}: The image of this observable is very similar to that of $\widehat{\cal K}_{2ss}(q^2)/\alpha_{\Lambda_c}$. For the sake of brevity, we will not repeat it here.

\item \underline{$\widehat{\cal K}_{2c}(q^2)/\alpha_{\Lambda_c}$}: The two best-fit points in the NP hypothesis $\left(C_{S_R},\, C_{S_L}\right)$ can also be distinguished by this observable clearly. The predicted value of this observable increases greatly in BP4B, but decreases slightly in BP4A. There are overlapping parts in the regions of scenarios BP1, BP2, and BP7. In scenarios BP1 and BP6, the predicted values of $\widehat{\cal K}_{2c}(q^2)/\alpha_{\Lambda_c}$ increase.

\item \underline{$\widehat{\cal K}_{3sc}(q^2)/\alpha_{\Lambda_c}$}: Consistent with our previous discussion in subsection~\ref{subsec:angular}, only scenarios BP2 and BP6 can provide nonzero $\widehat{\cal K}_{3sc}(q^2)$. This observable can distinguish between the scenario and its complex conjugate partner since the relationship $\widehat{\cal K}_{3sc}(q^2)\big|_{\text{BP}^*} = - \widehat{\cal K}_{3sc}(q^2)\big|_{\text{BP}}$, where BP stands for BP2 and BP6, holds. The $\widehat{\cal K}_{3sc}(q^2_{\rm max}) = 0$ is due to the disappearance of the transversity amplitudes ${\cal A}_{\parallel_t}$, ${\cal A}_{\perp_{1,0}}$, and ${\cal A}^T_{\perp_{1,0}}$ at the endpoint $q^2_{\rm max}$ and the relationships $G_+(q^2_{\rm max}) = G_\perp(q^2_{\rm max})$ and $\widetilde{h}_+(q^2_{\rm max}) = \widetilde{h}_\perp(q^2_{\rm max})$. The Eq.~\eqref{eq:K3sc} contains only the dimensionless factors $S_3$, $S^T_3$, and $R_3$. After integrating over $E_\pi$, they become $\bar{S}_3$, $\bar{S}^T_3$, and $\bar{R}_3$, respectively. Obviously, $\widehat{\cal K}_{3sc}(q^2_{\rm min}) = 0$ since $\bar{S}_3,\, \bar{S}^T_3,\, \bar{R}_3 \sim {\cal O}(\varepsilon^3)$.

\item \underline{$\widehat{\cal K}_{3s}(q^2)/\alpha_{\Lambda_c}$}: Consistent with our expectation in subsection~\ref{subsec:angular}, the predicted value of this observable is nonzero only in the three scenarios with complex phases. Since the relationship $\widehat{\cal K}_{3s}(q^2)\big|_{\text{BP}^*} = - \widehat{\cal K}_{3s}(q^2)\big|_{\text{BP}}$, where BP represents BP1, BP2, and BP6, holds, the scenario and its complex conjugate partner can be distinguished by $\widehat{\cal K}_{3s}(q^2)$. This observable can also clearly distinguish the scenarios BP1 and BP2. The transversity amplitudes ${\cal A}_{\parallel_t}$, ${\cal A}_{\perp_{1,0}}$, and ${\cal A}^T_{\perp_{1,0}}$ are vanishing at the endpoint $q^2_{\rm max}$ causing $\widehat{\cal K}_{3s}(q^2_{\rm max}) = 0$. 

\item \underline{$\widehat{\cal K}_{4sc}(q^2)/\alpha_{\Lambda_c}$}: Only the predicted value in scenario BP1 has a large deviation from that in the SM. Other NP scenarios are difficult to distinguish from the SM. The two endpoints of this observable are fixed at zero, since the Eq.~\eqref{eq:K4sc} contains only the dimensionless factors $S_3$, $S^T_3$, and $R_3$, and the transversity amplitudes ${\cal A}_{\parallel_t}$, ${\cal A}_{\perp_{1,0}}$, and ${\cal A}^T_{\perp_{1,0}}$ are vanishing at $q^2_{\rm max}$.

\item \underline{$\widehat{\cal K}_{4s}(q^2)/\alpha_{\Lambda_c}$}: In scenario BP4B, the predicted value of $\widehat{\cal K}_{4s}(q^2)/\alpha_{\Lambda_c}$ is a relatively large positive number (also see Table~\ref{tab:numres}), which is quite different from the predicted value in other NP scenarios and the SM. In scenarios BP1 and BP6, the predicted value of this observable is slightly improved.
\end{itemize}

Observables $\widehat{\cal K}_{3sc}(q^2)$ and $\widehat{\cal K}_{4sc}(q^2)$ only include the suppressed dimensionless factors $\bar{S}_3$, $\bar{S}^T_3$, and $\bar{R}_3$, making them an order of magnitude smaller than other observables. Except the $\widehat{\cal K}_{3sc}(q^2)$ and $\widehat{\cal K}_{3s}(q^2)$, other observables are not sensitive to the sign of imaginary part, and their predicted values in complex conjugate scenario BP$^*$ are exactly the same as those in scenario BP, where BP stands for BP1, BP2, and BP6. The NP effects in scenarios BP3 and BP5 are mainly the contribution of $C_{V_L}$, so they have little impact on the normalized angular observables $\widehat{{\cal K}}_i(q^2)$. There is always an overlap between the SM predictions and the predictions in scenario BP7. The values of the corresponding angular observables $\widehat{\cal K}_i$ are provided in Table~\ref{tab:numres}.

\section{Conclusions}
\label{sec:conclusion}

Inspired by the anomalies in $\bar{B} \to D^{(*)} \tau^- \bar{\nu}_\tau$ decays, many works have been done to explore possible NP patterns in $b \to c \tau^- \bar{\nu}_\tau$ transition by studying the baryonic counterparts, that is, the $\Lambda_b^0 \to \Lambda_c^+ \tau^- \bar{\nu}_\tau$ decay or the cascade decay $\Lambda_b^0 \to \Lambda_c^+ (\to \Lambda^0 \pi^+) \tau^- \bar{\nu}_\tau $. Comparing with $\bar{B} \to D^{(*)} \tau^- \bar{\nu}_\tau$ decays, the baryonic counterparts could be useful to confirm more possible Lorentz structures of the NP effective operators. However, the angular distribution of them cannot be measured since the solid angle of the final-state particle $\tau^-$ cannot be determined precisely. Therefore, in this work, we further consider the subsequent decay $\tau^- \to \pi^- \nu_\tau$  to construct a measurable angular distribution. The full process is $\Lambda_b^0 \to \Lambda_c^+ (\to \Lambda^0 \pi^+)\tau^- (\to \pi^- \nu_\tau)\bar{\nu}_\tau$, which includes three visible final-state particles $\Lambda^0$, $\pi^+$, and $\pi^-$ whose three momenta can be measured.

For an unpolarized initial $\Lambda_b$ state, the five-fold differential angular distribution including all Lorentz structures of the NP effective operators can be expressed in terms of ten angular observables ${\cal K}_i (q^2, E_\pi)$, which can be completely expressed by ten independent transversity amplitudes, the asymmetry parameter $\alpha_{\Lambda_c}$ related to $\Lambda_c^+ \to \Lambda^0 \pi^+$ decay, and some dimensionless factors given in Eqs.~\eqref{eq:St}--\eqref{eq:R3}. Our results are consistent with the Ward-like relation, and when the transversity states $\perp$ and $\parallel$ are exchanged, they have good symmetry or antisymmetry. We  also find that our results of $d\Gamma/dq^2$, which can be obtained by integrating over the kinematic parameters $E_\pi$, $\cos\theta_\Lambda$, $\cos\theta_\pi$, and $\phi_\pi$, are complete agreement with those in Ref.~\cite{Datta:2017aue}. Based on these, we believe that our results are correct.

If the angular distribution is found to contain the nonzero component $\sin\theta_\pi \cos\theta_\pi \sin\theta_\Lambda \sin\phi_\pi$, this will be an unquestionable sign of the NP, indicating that the tensor operator must exist and that the corresponding Wilson coefficient $g_T$ has a different weak phase than $g_V$ or $g_A$.

We obtain a number of observables by integrating over some of the five kinematic parameters. On the hadron side, there are the $\Lambda_c$ spin polarization $P_{\Lambda_c}(q^2)$ and certainly the differential decay rate $d\Gamma/dq^2$. Since all the lepton-side kinematic parameters have been integrated over, these observables are not affected by $\tau^-$ decay dynamics, so their expressions are applicable to light leptons $\ell = \mu,\,e$ (Necessary replacement $m_\tau \to m_\ell$ and removal of factor ${\cal B}(\tau \to \pi^- \nu_\tau)$ are required). On the lepton side, there are the three-fold differential angular distributions $d^3\Gamma/(dq^2 dE_\pi d\cos\theta_\pi)$ and $d^3\Gamma/(dq^2 dE_\pi d\phi_\pi)$, and the two-fold differential decay rate $d^2\Gamma/(dq^2 dE_\pi)$, as well as the $\pi^-$ meson forward-backward asymmetry $A_{FB}(q^2)$. These observables depend on at least one kinematic parameter of $\pi^-$, so they only exist in $\tau$ channels, and specifically for the $\tau \to \pi^- \nu_\tau$ decay. The $P_{\Lambda_c}(q^2)$ and $A_{FB}(q^2)$ can be represented by the angular observables $\widehat{{\cal K}}_i (q^2)$.

Using the $\Lambda_b \to \Lambda_c$ form factors computed in lattice QCD including all the types of Lorentz structures of the NP effective operators, we predict the entire set of the angular observables $\widehat{{\cal K}}_i (q^2)$ and $\widehat{{\cal K}}_i$ both within the SM and in some NP scenarios, which are a variety of best-fit solutions in seven different NP hypotheses. We find that the two best-fit points in the NP hypothesis $\left(C_{S_R},\, C_{S_L}\right)$, which is motivated by models with extra charged Higgs, can be distinguished by observables $\widehat{\cal K}_{1c}$, $\widehat{\cal K}_{2ss}$, $\widehat{\cal K}_{2cc}$, $\widehat{\cal K}_{2c}$, and $\widehat{\cal K}_{4s}$. Although the uncertainties of the NP parameters and the $\Lambda_b\to \Lambda_c$ form factors are taken into account, the predicted values in scenario BP1 are still accurate. This allows scenario BP1 to be well distinguished from other scenarios and the SM. The predicted values of observables $\widehat{\cal K}_{1c}$, $\widehat{\cal K}_{2ss}$, $\widehat{\cal K}_{2cc}$, $\widehat{\cal K}_{3sc}$, and $\widehat{\cal K}_{3s}$ in scenario BP2 are significantly different from the predicted values in the SM.

The (HL-)LHC will produce a large number of $\Lambda_b$ baryons, with a production cross section $\sigma(\Lambda_b)/\sigma(b\bar{b}) \sim 10\%$. Future precise measurements of the angular observables in $\Lambda_b^0 \to \Lambda_c^+ (\to \Lambda^0 \pi^+)\tau^- (\to \pi^- \nu_\tau)\bar{\nu}_\tau$ decay, especially precise measurements of the normalized ones, would be very helpful to provide a more definite answer concerning the observed anomalies by the BaBar, Belle, and LHCb collaborations, restricting further or even deciphering the NP models.

\section*{Acknowledgements}

This work is supported by the National Natural Science Foundation of China under Grant Nos.~11947083, 12075097, 11675061 and 11775092.

\appendix

\section{The detailed calculation of the measurable angular distribution}
\label{sec:app}

The differential decay rate of the unpolarized $\Lambda_b^0 \to \Lambda_c^+ (\to \Lambda^0 \pi^+)\tau^- (\to \pi^- \nu_\tau)\bar{\nu}_\tau$ decay can be written as
\begin{equation}
\label{eq:DDR11}
d\Gamma = \frac{1}{2m_{\Lambda_b}}\left|{\cal M} \right|^2 d\Pi_5(p_{\Lambda_b};p_{\pi^-},p_\nu,p_{\bar{\nu}},p_\Lambda,p_{\pi^+}),  
\end{equation}
where the squared matrix element is 
\begin{align}
&\left|{\cal M} \right|^2 = \sum_{\lambda_\Lambda}\frac{1}{2}\sum_{\lambda_{\Lambda_b}}\left| {\cal M}_{\lambda_{\Lambda_b}}^{\lambda_\Lambda}\right|^2 ,
\nonumber \\[2mm]
&{\cal M}_{\lambda_{\Lambda_b}}^{\lambda_\Lambda} =  \sum_{\lambda_{\Lambda_c},\lambda_\tau}
\frac{{\cal M}_{\lambda_{\Lambda_b}}^{\lambda_{\Lambda_c},\lambda_\tau}(\Lambda_b\to \Lambda_c \tau \bar{\nu}_\tau) {\cal M}_{\lambda_{\Lambda_c}}^{\lambda_\Lambda}(\Lambda_c\to \Lambda \pi^+) {\cal M}_{\lambda_\tau}(\tau\to \pi^- \nu_\tau)}
{(p^2_{\Lambda_c} - m^2_{\Lambda_c} + i m_{\Lambda_c} \Gamma_{\Lambda_c})(p^2_{\tau} - m^2_{\tau} + i m_{\tau} \Gamma_{\tau})},
\end{align}
as well as the five-body phase space\footnote{In this work, the $n$-body phase space is most generally defined as 
\begin{equation*}
d\Pi_n(P;p_i) = \left( \prod_{i}\frac{d^3 p_i}{(2\pi)^3 2E_i} \right) (2\pi)^4 \delta^{(4)}\left( P - \sum p_i\right). 
\end{equation*}} is 
\begin{align}
d\Pi_5(p_{\Lambda_b};p_{\pi^-},p_\nu,p_{\bar{\nu}},p_\Lambda,p_{\pi^+}) = 
&\frac{dq^2 dp^2_{\tau} dp^2_{\Lambda_c}}{(2\pi)^3}
d\Pi_2(p_{\Lambda_b};q,p_{\Lambda_c})
\nonumber \\[2mm]
& \times d\Pi_2(q;p_{\tau},p_{\bar{\nu}}) 
d\Pi_2(p_{\tau};p_{\pi^-},p_\nu) d\Pi_2(p_{\Lambda_c};p_\Lambda,p_{\pi^+}).
\label{eq:phase}
\end{align}
The $\lambda_x$ stands for the helicity of the particle $x$. We drop the helicity indices $\lambda_{\pi^\pm}$ as they are null and fix $\lambda_{\bar{\nu}_\tau}$ ($\lambda_{\nu_\tau}$) to $\frac{1}{2}$ ($-\frac{1}{2}$). 

Using the effective Hamiltonian given in Eq.~\eqref{eq:Heff}, one can express the helicity amplitude of $\Lambda_b\to \Lambda_c \tau \bar{\nu}_\tau$ decay as 
\begin{equation}
\label{eq:3bodyHA}
{\cal M}_{\lambda_{\Lambda_b}}^{\lambda_{\Lambda_c},\lambda_\tau}(\Lambda_b\to \Lambda_c \tau \bar{\nu}_\tau) = \sqrt{2}G_F V_{cb} \left(  H_{\lambda_{\Lambda_b}}^{\lambda_{\Lambda_c}} L^{\lambda_\tau} + \sum_{\lambda}\eta_{\lambda} H^{\lambda_{\Lambda_c},\lambda}_{\lambda_{\Lambda_b}} L^{\lambda_\tau}_{\lambda} +\sum_{\lambda,\lambda'} \eta_{\lambda} \eta_{\lambda'} H^{\lambda_{\Lambda_c},\lambda,\lambda'}_{\lambda_{\Lambda_b}} L^{\lambda_\tau}_{\lambda,\lambda'}\right).
\end{equation}
Here $\lambda^{(\prime)} = t, \,\pm1, \, 0$ indicates the helicity of the virtual vector boson $W^{*}$. The number of the helicity indexes depends on the Lorentz structure of the effective operator. The factor $\eta$ that appears here is due to the use of the completeness relation (Eq.~\eqref{eq:completeness_relation}) of the polarization vectors of the virtual vector boson. The hadronic and leptonic helicity amplitudes are respectively defined as
\begin{align}
&H^{\lambda_{\Lambda_c}}_{\lambda_{\Lambda_b}} \equiv \left\langle \Lambda_c(\lambda_{\Lambda_c}) \left|g_S ({\bar c} b) + g_P ({\bar c} \gamma_5 b) \right|\Lambda_b(\lambda_{\Lambda_b}) \right\rangle, 
\\[2mm]
&H^{\lambda_{\Lambda_c},\lambda}_{\lambda_{\Lambda_b}} \equiv \epsilon^{\mu*}(\lambda)\left\langle \Lambda_c(\lambda_{\Lambda_c}) \left|g_V ({\bar c}\gamma_\mu b) + g_A ({\bar c}\gamma_\mu \gamma_5 b) \right|\Lambda_b(\lambda_{\Lambda_b}) \right\rangle, 
\\[2mm]
&H^{\lambda_{\Lambda_c},\lambda,\lambda'}_{\lambda_{\Lambda_b}} \equiv g_T \epsilon^{\mu*}(\lambda) \epsilon^{\nu*}(\lambda') \left\langle \Lambda_c(\lambda_{\Lambda_c}) \left| {\bar c} i \sigma_{\mu\nu}(1-\gamma_5) b \right|\Lambda_b(\lambda_{\Lambda_b}) \right\rangle, 
\end{align}
and
\begin{align}
&L^{\lambda_\tau} \equiv \left\langle \tau^-(\lambda_\tau)\bar{\nu}\left| {\bar\tau} P_L\nu\right| 0\right\rangle, \\[2mm]
&L^{\lambda_\tau}_{\lambda} \equiv \epsilon^{\mu}(\lambda)\left\langle \tau^-(\lambda_\tau)\bar{\nu}\left| {\bar\tau}\gamma_\mu P_L\nu\right| 0\right\rangle, 
\\[2mm]
&L^{\lambda_\tau}_{\lambda,\lambda'}\equiv 
(-i)\epsilon^{\mu}(\lambda) \epsilon^{\nu}(\lambda')\left\langle \tau^-(\lambda_\tau)\bar{\nu}\left| {\bar\tau}\sigma_{\mu\nu} P_L\nu\right| 0\right\rangle, 
\end{align}
where $\epsilon^\mu(\lambda)$ is the polarization vector of the virtual vector boson with helicity $\lambda$.  

Using the narrow width ($\Gamma_y \ll m_y$) approximation
\begin{equation}
\frac{1}{(p^2_y - m^2_y)^2 + m^2_y \Gamma^2_y} = \frac{\pi}{m_y \Gamma_y} \delta(p^2_y - m^2_y),\quad (y=\Lambda_c,\,\tau)
\end{equation}
in Eq.~\eqref{eq:DDR11} and by integrating over the $dp^2_\tau dp^2_{\Lambda_c}$, one can obtain two on-shell relations $p^2_{\Lambda_c} = m^2_{\Lambda_c}$ and $p^2_\tau = m^2_\tau$, as well as
\begin{align}
d\Gamma =& \frac{dq^2}{2^5 \pi m_{\Lambda_b} m_\tau \Gamma_\tau m_{\Lambda_c} \Gamma_{\Lambda_c}} 
d\Pi_2(p_{\Lambda_b};q,p_{\Lambda_c})
d\Pi_2(q;p_{\tau},p_{\bar{\nu}}) 
d\Pi_2(p_{\tau};p_{\pi^-},p_\nu) d\Pi_2(p_{\Lambda_c};p_\Lambda,p_{\pi^+})
\nonumber \\[2mm]
&\times
\sum_{\lambda_\Lambda,\lambda_{\Lambda_b}}\left| \sum_{\lambda_{\Lambda_c},\lambda_\tau} {\cal M}_{\lambda_{\Lambda_b}}^{\lambda_{\Lambda_c},\lambda_\tau}(\Lambda_b\to \Lambda_c \tau \bar{\nu}_\tau) {\cal M}_{\lambda_{\Lambda_c}}^{\lambda_\Lambda}(\Lambda_c\to \Lambda \pi^+) {\cal M}_{\lambda_\tau}(\tau\to \pi^- \nu_\tau)
\right|^2.
\label{eq:dG2}
\end{align}
Since each individual two-body phase space or helicity amplitude is Lorentz invariant in Eqs.~\eqref{eq:dG2} and \eqref{eq:3bodyHA}, one can finish each part of $d\Gamma$ in different reference frames. In this work, we should consider three measurable reference frames --- the $\Lambda_b$ rest frame, the $\Lambda_c$ rest frame and the $\tau^-\bar{\nu}_\tau$ center-of-mass frame.

\subsection{In the $\Lambda_b$ rest frame}
\label{subsec:Lambdab_RF}

In this frame, we calculate the hadronic helicity amplitudes $H$ and the two-body phase space $d\Pi_2(p_{\Lambda_b};q,p_{\Lambda_c})$. We choose the three-momentum of the $\Lambda_c$ baryon to point to the $+z$ direction and the three-momentum of the virtual vector boson $W^{*}$ to point to the $-z$ direction, see Figure~\ref{fig:angle}. The momenta of $\Lambda_b$, $\Lambda_c$, and $W^{*}$ are respectively given by 
\begin{align}
p_{\Lambda_b}^{\mu} = (m_{\Lambda_b},0,0,0), 
\qquad
p_{\Lambda_c}^{\mu} = (E_{\Lambda_c},0,0,\left| {\bm p}_{\Lambda_c}\right| ), 
\qquad
q^\mu = (q_0, 0, 0, -\left| {\bm q}\right| ). 
\end{align}
The spinors of $\Lambda_b$ and $\Lambda_c$ are then given by~\cite{Auvil:1966eao,Haber:1994pe}
\begin{align}
&u_{\Lambda_b}\left( \frac{1}{2}\right)  = \left( \sqrt{2m_{\Lambda_b}},0,0,0\right)^T,
&&u_{\Lambda_b}\left( -\frac{1}{2}\right)  = \left( 0,\sqrt{2m_{\Lambda_b}},0,0\right)^T,
\\[2mm]
&u_{\Lambda_c}\left( \frac{1}{2}\right)  = \left( \beta^+_{\Lambda_c},0,\beta^-_{\Lambda_c},0\right)^T,
&&u_{\Lambda_c}\left( -\frac{1}{2}\right)  = \left( 0,\beta^+_{\Lambda_c},0,-\beta^-_{\Lambda_c}\right)^T,
\end{align}
with $\beta^\pm_x \equiv \sqrt{E_x \pm m_x}$. In this frame, the polarization vectors of the virtual vector boson $W^{*}$ can be written as~\cite{Auvil:1966eao,Haber:1994pe}
\begin{equation}
\epsilon^\mu(t) = q^\mu/\sqrt{q^2},
\end{equation}
corresponding to $J_W=0,\lambda_W=0$, and
\begin{align}
\epsilon^\mu(\pm 1) &= (0, \pm 1, -i, 0)/\sqrt{2},
\\[2mm]
\epsilon^\mu(0) &= (|{\bm q}|,0,0,-q_0)/\sqrt{q^2},
\end{align}
corresponding to $J_W=1,\lambda_W= \pm 1 ,0$. The well-known completeness relation can be expressed as
\begin{equation}
\label{eq:completeness_relation}
g^{\mu\nu}=\sum_{\lambda \in \{t,\pm1,0\}} \epsilon^{\mu}(\lambda) \epsilon^{\nu*}(\lambda) \eta_{\lambda},
\end{equation}
with $\eta_{t} = 1$ and $\eta_{\pm1,0} = -1$.

By integrating over the two-body phase space, we can get
\begin{equation}
\int d\Pi_2(p_{\Lambda_b};q,p_{\Lambda_c}) = \frac{\left|{\bm p}_{\Lambda_c} \right| }{4\pi m_{\Lambda_b}},
\end{equation}
as well as $\left|{\bm p}_{\Lambda_c} \right|=\left|{\bm q} \right|= \lambda^{1/2}(m^2_{\Lambda_b}, m^2_{\Lambda_c}, q^2)/(2m_{\Lambda_b})$, $E_{\Lambda_c} = (m_{\Lambda_b}^2 + m_{\Lambda_c}^2 - q^2) / (2 m_{\Lambda_b})$, and $q_0 = (m_{\Lambda_b}^2 - m_{\Lambda_c}^2 + q^2) / (2 m_{\Lambda_b})$. The K{\"a}ll\'en function $\lambda(a,b,c) \equiv a^2 + b^2 + c^2 -2ab - 2ac - 2bc$, and $\lambda(m^2_{\Lambda_b}, m^2_{\Lambda_c}, q^2) = Q_+ Q_-$.

The nonzero hadronic helicity amplitudes $H$ can be expressed by the transversity amplitudes as follows
\begin{align}
H^{1/2}_{1/2} &= \left({\cal A}^{SP}_{\perp_t} + {\cal A}^{SP}_{\parallel_t} \right) /\sqrt{2},
&
H^{-1/2}_{-1/2} &= \left({\cal A}^{SP}_{\perp_t} - {\cal A}^{SP}_{\parallel_t} \right) /\sqrt{2},
\\[2mm]
H^{1/2,t}_{1/2} &= \left({\cal A}^{VA}_{\perp_t} + {\cal A}^{VA}_{\parallel_t} \right) /\sqrt{2},
&
H^{-1/2,t}_{-1/2} &= \left({\cal A}^{VA}_{\perp_t} - {\cal A}^{VA}_{\parallel_t} \right) /\sqrt{2},
\\[2mm]
H^{1/2,1}_{-1/2} &= \left({\cal A}_{\perp_1} + {\cal A}_{\parallel_1} \right) /\sqrt{2},
&
H^{-1/2,-1}_{1/2} &= \left({\cal A}_{\perp_1} - {\cal A}_{\parallel_1} \right) /\sqrt{2},
\\[2mm]
H^{1/2,0}_{1/2} &= \left({\cal A}_{\perp_0} + {\cal A}_{\parallel_0} \right) /\sqrt{2},
&
H^{-1/2,0}_{-1/2} &= \left({\cal A}_{\perp_0} - {\cal A}_{\parallel_0} \right) /\sqrt{2},
\\[2mm]
H^{-1/2,-1,t}_{1/2} &= H^{-1/2,0,-1}_{1/2} = \frac{{\cal A}^T_{\perp_1} - {\cal A}^T_{\parallel_1} } { 2\sqrt{2}} ,
&
H^{1/2,1,-1}_{1/2} &= H^{1/2,0,t}_{1/2} = \frac{{\cal A}^T_{\perp_0} + {\cal A}^T_{\parallel_0}} {2\sqrt{2}} ,
\\[2mm]
H^{1/2,1,0}_{-1/2} &= H^{1/2,1,t}_{-1/2} = \frac{{\cal A}^T_{\perp_1} + {\cal A}^T_{\parallel_1} }{ 2\sqrt{2}} ,
&
H^{-1/2,1,-1}_{-1/2} &= H^{-1/2,0,t}_{-1/2} = \frac{{\cal A}^T_{\perp_0} - {\cal A}^T_{\parallel_0}}{ 2\sqrt{2}} ,
\end{align}
together with the other eight non-vanishing tensor-type helicity amplitudes related to the above ones by
\begin{equation}
H^{\lambda_{\Lambda_c},\lambda,\lambda'}_{\lambda_{\Lambda_b}} = - H^{\lambda_{\Lambda_c},\lambda',\lambda}_{\lambda_{\Lambda_b}}.
\end{equation}

\subsection{In the $\Lambda_c$ rest frame}
\label{subsec:Lambdac_RF}

In this frame, we calculate the helicity amplitude ${\cal M}_{\lambda_{\Lambda_c}}^{\lambda_\Lambda}(\Lambda_c\to \Lambda \pi^+)$ and the two-body phase space $d\Pi_2(p_{\Lambda_c};p_\Lambda,p_{\pi^+})$. The momenta of $\Lambda_c$ and $\Lambda$ are respectively given by
\begin{equation}
\tilde{p}^\mu_{\Lambda_c} = (m_{\Lambda_c},0,0,0), 
\quad
p^\mu_{\Lambda} = (E_\Lambda, \left| {\bm p}_{\Lambda}\right|\sin\theta_\Lambda, 0, \left| {\bm p}_{\Lambda}\right|\cos\theta_\Lambda).
\end{equation}
The `` $\tilde{\;}$ " here and the following are only used to distinguish the representations of the same kinematic quantity in different reference frames. The spinors of $\Lambda_c$ and $\Lambda$ are given by~\cite{Auvil:1966eao,Haber:1994pe}
\begin{align}
&\tilde{u}_{\Lambda_c}\left( \frac{1}{2}\right)  = \left( \sqrt{2m_{\Lambda_c}},0,0,0\right) ^T,
\qquad
\tilde{u}_{\Lambda_c}\left( -\frac{1}{2}\right)  = \left( 0,\sqrt{2m_{\Lambda_c}},0,0\right) ^T,
\\[2mm]
&u_{\Lambda}\left( \frac{1}{2}\right)  = \left( \beta^+_{\Lambda}\cos\frac{\theta_\Lambda}{2},\beta^+_{\Lambda}\sin\frac{\theta_\Lambda}{2},\beta^-_{\Lambda}\cos\frac{\theta_\Lambda}{2},\beta^-_{\Lambda}\sin\frac{\theta_\Lambda}{2}\right) ^T,
\\[2mm]
&u_{\Lambda}\left( -\frac{1}{2}\right)  = \left( -\beta^+_{\Lambda}\sin\frac{\theta_\Lambda}{2},\beta^+_{\Lambda}\cos\frac{\theta_\Lambda}{2},\beta^-_{\Lambda}\sin\frac{\theta_\Lambda}{2},-\beta^-_{\Lambda}\cos\frac{\theta_\Lambda}{2}\right) ^T,
\end{align}

By integrating over the $\delta^{(4)}$ term and the azimuthal angle $\phi_\Lambda$ in two-body phase space, one can get
\begin{equation}
d\Pi_2(p_{\Lambda_c};p_\Lambda,p_{\pi^+}) = \frac{1}{8\pi}\frac{\left|{\bm p}_\Lambda \right| }{m_{\Lambda_c}} d\cos\theta_{\Lambda},
\end{equation}
as well as $\left| {\bm p}_\Lambda\right| = \lambda^{1/2}(m^2_{\Lambda_c}, m^2_\Lambda,m^2_\pi)/(2m_{\Lambda_c})$ and $E_\Lambda = (m^2_{\Lambda_c} + m^2_\Lambda - m^2_\pi)/(2m_{\Lambda_c})$.

The helicity amplitude
\begin{align}
{\cal M}_{\lambda_{\Lambda_c}}^{\lambda_\Lambda}(\Lambda_c\to \Lambda \pi^+) &= i\bar{u}_\Lambda(\lambda_\Lambda)(A+B\gamma_5)u_{\Lambda_c}(\lambda_{\Lambda_c}),
\nonumber\\[2mm]
&=i\chi^\dagger_\Lambda(\lambda_\Lambda)\left( S + P {\bm \sigma}\cdot \hat{\bm p}_\Lambda\right) \chi_{\Lambda_c}(\lambda_{\Lambda_c}).
\end{align}
Where $S=\sqrt{2m_{\Lambda_c}}\beta^+_\Lambda A$ stands for the parity-violating $s$-wave amplitude and $P=-\sqrt{2m_{\Lambda_c}}\beta^-_\Lambda B$ stands for the parity-conserving $p$-wave amplitude. ${\bm \sigma} = (\sigma^1,\sigma^2,\sigma^3)$ is a vector composed of Pauli matrices. $\hat{\bm p}_\Lambda$ is the unit vector along the direction of $\Lambda$ baryon. The four helicity amplitudes are
\begin{align}
&{\cal M}^{1/2}_{1/2}(\Lambda_c\to \Lambda \pi^+) = i(S+P)\cos\frac{\theta_\Lambda}{2},
&&{\cal M}^{1/2}_{-1/2}(\Lambda_c\to \Lambda \pi^+) = i(S+P)\sin\frac{\theta_\Lambda}{2},
\label{eq:LambdacHA1}
\\[2mm]
&{\cal M}^{-1/2}_{1/2}(\Lambda_c\to \Lambda \pi^+) = -i(S-P)\sin\frac{\theta_\Lambda}{2},
&&{\cal M}^{-1/2}_{-1/2}(\Lambda_c\to \Lambda \pi^+) = i(S-P)\cos\frac{\theta_\Lambda}{2}.
\label{eq:LambdacHA2}
\end{align}
The two helicity amplitudes in Eq.~\eqref{eq:LambdacHA1} correspond to $\lambda_\Lambda = \frac{1}{2}$. By using them, one can obtain that the decay rate of $\Lambda_c\to \Lambda \pi^+$ with $\lambda_\Lambda = \frac{1}{2}$ is
\begin{equation}
\Gamma^{\lambda_\Lambda=1/2}(\Lambda_c\to \Lambda \pi^+) =\frac{\left| {\bm p}_\Lambda\right| }{16\pi m^2_{\Lambda_c}}\left|S+P \right|^2.
\end{equation}
In the same way, one can obtain the decay rate
\begin{equation}
\Gamma^{\lambda_\Lambda=-1/2}(\Lambda_c\to \Lambda \pi^+) =\frac{\left| {\bm p}_\Lambda\right| }{16\pi m^2_{\Lambda_c}}\left|S-P \right|^2, 
\end{equation}
by using the two helicity amplitudes in Eq.~\eqref{eq:LambdacHA2}. The angular asymmetry parameter $\alpha_{\Lambda_c}$ is defined as
\begin{equation}
\label{eq:alphaLambdac}
\alpha_{\Lambda_c} \equiv \frac{2\Re(S^*P)}{\left|S\right|^2 +\left| P \right|^2},
\end{equation}
and one can immediately get the relations
\begin{equation}
\frac{\Gamma^{\lambda_\Lambda = 1/2}}{\Gamma^{\lambda_\Lambda = 1/2} + \Gamma^{\lambda_\Lambda = -1/2}} = \frac{1}{2}(1+\alpha_{\Lambda_c}),
\quad
\frac{\Gamma^{\lambda_\Lambda = -1/2}}{\Gamma^{\lambda_\Lambda = 1/2} + \Gamma^{\lambda_\Lambda = -1/2}} = \frac{1}{2}(1-\alpha_{\Lambda_c}).
\end{equation}

\subsection{In the $\tau^-\bar{\nu}_\tau$ center-of-mass frame}
\label{subsec:dileptonCMS}

In this frame, we calculate the leptonic helicity amplitudes $L$ and the helicity amplitudes ${\cal M}_{\lambda_\tau}(\tau\to \pi^- \nu_\tau)$, as well as the two-body phase spaces $d\Pi_2(q;p_{\tau},p_{\bar{\nu}})$ and $d\Pi_2(p_{\tau};p_{\pi^-},p_\nu)$. The momentum of $\pi^-$ is defined as $p^\mu_{\pi} = (E_\pi, \left| {\bm p}_\pi\right|\hat{\bm p}_\pi )$ with
\begin{equation}
\hat{\bm p}_\pi=(\sin\theta_\pi \cos\phi_\pi, \sin\theta_\pi \sin\phi_\pi, \cos\theta_\pi),
\end{equation} 
is the unit vector along the direction of $\pi^-$. In this frame, the polarization vectors of the virtual vector boson $W^{*}$ are changed to
\begin{equation}
\tilde{\epsilon}^\mu(t)=(1,0,0,0),
\quad
\tilde{\epsilon}^\mu(\pm1)=(0,\pm1,-i,0)/\sqrt{2},
\quad
\tilde{\epsilon}^\mu(0)=(0,0,0,-1). 
\end{equation}

The helicity amplitudes ${\cal M}_{\lambda_\tau}(\tau\to \pi^- \nu_\tau)$ can be written as
\begin{equation}
{\cal M}_{\lambda_\tau}(\tau\to \pi^- \nu_\tau) = i\sqrt{2}G_F V_{ud}^* f_\pi \bar{u}_{\nu_\tau} {\slashed p}_\pi P_L u_\tau(\lambda_\tau),
\label{eq:taudecayM}
\end{equation}
and one can obtain that the decay rate is
\begin{equation}
\Gamma(\tau \to \pi^- \nu_\tau)
= \frac{G_F^2\left|V_{ud} \right|^2 f_\pi^2 (m_\tau^2 - m_\pi^2)^2}{16 \pi m_\tau}.
\end{equation}

Using the relation $\sum_{\lambda_\tau} u_\tau(\lambda_\tau) \bar{u}_\tau(\lambda_\tau) = {\slashed p}_\tau + m_\tau$, we link the $\bar{u}_{\nu_\tau} {\slashed p}_\pi P_L u_\tau(\lambda_\tau)$ in Eq.~\eqref{eq:taudecayM} with the leptonic helicity amplitudes $L$, and we can obtain the new leptonic helicity amplitudes as follows
\begin{align}
&{\sf L} 
= m_\tau p^\mu_\pi \bar{u}_{\nu_\tau} \gamma_\mu P_L v_{\bar{\nu}_\tau},
\\[2mm]
&{\sf L}_\lambda 
= m_\tau^2 \epsilon^\mu(\lambda) \bar{u}_{\nu_\tau} \gamma_\mu P_L v_{\bar{\nu}_\tau},
\\[2mm]
&{\sf L}_{\lambda,\lambda'} 
=m_\tau \left[ p_\pi \cdot \epsilon(\lambda) \epsilon^\mu(\lambda') - p_\pi \cdot \epsilon(\lambda') \epsilon^\mu(\lambda) + i\epsilon^{\rho \alpha \nu \mu} p_{\pi \rho} \epsilon_\alpha(\lambda) \epsilon_\nu(\lambda')\right] \bar{u}_{\nu_\tau} \gamma_\mu P_L v_{\bar{\nu}_\tau}.
\end{align}

Next, we deduce the phase spaces $d\Pi_2(q;p_{\tau},p_{\bar{\nu}})$ and $d\Pi_2(p_{\tau};p_{\pi^-},p_\nu)$ simultaneously in the $\tau^-\bar{\nu}_\tau$ center-of-mass frame~\cite{Bhattacharya:2020lfm}.
\begin{align}
&d\Pi_2(q;p_{\tau},p_{\bar{\nu}})d\Pi_2(p_{\tau};p_{\pi},p_\nu)
\nonumber \\[2mm]
&=\int_{\delta^{(8)}} \frac{d^3 {\bm p}_{\tau}}{(2\pi)^3 2E_{\tau}}\frac{d^3 {\bm p}_{\bar{\nu}}}{(2\pi)^3 2E_{\bar{\nu}}} (2\pi)^4 \delta^{(4)}\left( q - p_{\tau} - p_{\bar{\nu}}\right) \frac{d^3 {\bm p}_{\pi}}{(2\pi)^3 2E_{\pi}}\frac{d^3 {\bm p}_\nu}{(2\pi)^3 2E_\nu} (2\pi)^4 \delta^{(4)}\left( p_\tau - p_{\pi} - p_\nu\right)
\nonumber \\[2mm]
&=\int_{\delta^{(2)}} \frac{1}{2^8 \pi^4} \frac{d^3 {\bm p}_{\tau}}{E_{\tau} \left|{\bm p}_\tau \right| } \delta\left( \sqrt{q^2} - E_{\tau} - \left|{\bm p}_\tau \right|\right) \frac{d^3 {\bm p}_{\pi}}{E_{\pi} \left| {\bm p}_\tau - {\bm p}_\pi \right| } \delta\left( E_\tau - E_{\pi} -\left| {\bm p}_\tau - {\bm p}_\pi \right|\right).
\end{align}
The momentum-conservation relations ${\bm p}_{\bar{\nu}} = -{\bm p}_\tau$ and ${\bm p}_\nu = {\bm p}_\tau - {\bm p}_\pi$ hold. Since three-momentum ${\bm p}_\pi$ can be measured experimentally, the remaining two $\delta$ functions will be used to integrate over the two variables in $d^3{\bm p}_\tau$. We define the solid angle of $\tau^-$ relative to the direction of $\pi^-$ instead of the $z$-axis as $(\theta_{\pi\tau}, \phi_{\pi\tau})$\footnote{The relationships between $(\theta_{\pi\tau}, \phi_{\pi\tau})$ and the solid angle of $\tau^-$ relative to $z$-axis $(\theta_{\tau}, \phi_{\tau})$, which can not be measured experimentally, are
\begin{align*}
&\cos\theta_{\pi\tau} = \cos\theta_\pi \cos\theta_\tau + \sin\theta_\pi \sin\theta_\tau \cos(\phi_\pi-\phi_\tau),
\\[2mm]
&\cos\phi_{\pi\tau} = \frac{\sin\theta_\tau \sin(\phi_\pi-\phi_\tau)}{\sqrt{\sin^2\theta_\tau \sin^2(\phi_\pi-\phi_\tau) + (\cos\theta_\tau \sin\theta_\pi - \cos\theta_\pi \sin\theta_\tau \cos(\phi_\pi - \phi_\tau))^2}},
\\[2mm]
&\sin\phi_{\pi\tau} = \frac{-\cos\theta_\tau \sin\theta_\pi + \cos\theta_\pi \sin\theta_\tau \cos(\phi_\pi - \phi_\tau)}{\sqrt{\sin^2\theta_\tau \sin^2(\phi_\pi-\phi_\tau) + (\cos\theta_\tau \sin\theta_\pi - \cos\theta_\pi \sin\theta_\tau \cos(\phi_\pi - \phi_\tau))^2}}.
\end{align*}}. Next, we will see that the magnitude $\left| {\bm p}_\tau\right| $ and the $\pi^- -\tau^-$ opening angle $\theta_{\pi\tau}$ can be determined theoretically.

Using formula $\delta(g(t)) = \sum_{i}\delta(t-t_i)/\left|g'(t_i) \right| $ where $g(t_i)=0$ and $g'(t_i) \neq 0$ to deduce the remaining two $\delta$ functions, one has
\begin{equation}
d\Pi_2(q;p_{\tau},p_{\bar{\nu}})d\Pi_2(p_{\tau};p_{\pi},p_\nu)
= \frac{1}{2^8\pi^4}\frac{1}{\sqrt{q^2}} d\phi_{\pi \tau} dE_\pi d\cos\theta_\pi d\phi_\pi, 
\end{equation}
as well as 
\begin{equation}
\left|{\bm p}_\tau \right| = \frac{q^2 - m_\tau^2}{2\sqrt{q^2}},
\quad
\cos\theta_{\pi\tau} = \frac{2 E_\tau E_\pi - m_\tau^2 - m_\pi^2}{2 \left|{\bm p}_\tau \right| \left|{\bm p}_\pi \right|}.
\end{equation}
Accordingly, the variables $q^2$ and $E_\pi$ can take
\begin{equation}
m_\tau^2 \leq q^2 \leq \left(m_{\Lambda_b} - m_{\Lambda_c} \right)^2, 
\quad
\frac{m_\tau^4 + m_\pi^2 q^2}{2 m_\tau^2 \sqrt{q^2}} \leq E_\pi \leq \frac{m_\pi^2 + q^2}{2 \sqrt{q^2}}.
\end{equation}
So far, all of the pieces of Eq.~\eqref{eq:dG2} have been completed.

\subsection{The five-fold differential decay rate}
\label{subsec:FFDW}

The five-fold differential decay rate is 
\begin{align}
\frac{d^5\Gamma}{dq^2 dE_\pi d\cos\theta_\pi d\phi_\pi d\cos\theta_\Lambda}
=& \frac{G_F^2 \left| V_{cb}\right|^2  \left| {\bm p}_{\Lambda_c}\right| (q^2)^{3/2} m_\tau^2}{2^8 \pi^4 m_{\Lambda_b}^2 (m_\tau^2 - m_\pi^2)^2} {\cal B}(\tau \to \pi^- \nu_\tau) {\cal B}(\Lambda_c \to \Lambda \pi^+)
\nonumber \\[2mm]
& \times \sum_{i,j}\left( {\cal N}^S_i |{\cal A}_i|^2 + {\cal N}^R_{i,j} {\rm Re}[{\cal A}_i {\cal A}_j^*] + {\cal N}^I_{i,j} {\rm Im}[{\cal A}_i {\cal A}_j^*]\right),
\label{eq:FFDW} 
\end{align}
where the terms ${\cal N}^S_i |{\cal A}_i|^2$, ${\cal N}^R_{i,j} {\rm Re}[{\cal A}_i {\cal A}_j^*]$, and ${\cal N}^I_{i,j} {\rm Im}[{\cal A}_i {\cal A}_j^*]$ are respectively listed in Table~\ref{tab:NS}, \ref{tab:NR}, and \ref{tab:NI}.

%%%%%%%%%%%%%%%%%%%%%%%%%%%%%%%%%%%%%%%%%%%%
\begin{table}
\begin{center}
\begin{tabular}{|c|c|} 
\hline
Transversity Amplitudes & ${\cal N}^S$ 
\\ \hline
$|{\cal A}_{\perp_t}|^2$ & $S_t$
\\ \hline
$|{\cal A}_{\parallel_t}|^2$ & $S_t$
\\ \hline
$|{\cal A}_{\perp_1}|^2$ & $S_1 + \alpha_{\Lambda_c} S_2 \cos\theta_{\Lambda} \cos\theta_\pi + S_3 \cos2\theta_\pi$
\\ \hline
$|{\cal A}_{\parallel_1}|^2$ & $S_1 + \alpha_{\Lambda_c} S_2 \cos\theta_{\Lambda} \cos\theta_\pi + S_3 \cos2\theta_\pi$
\\ \hline
$|{\cal A}_{\perp_0}|^2$ & $(S_1 - S_3) - 2 S_3 \cos2\theta_\pi$
\\ \hline
$|{\cal A}_{\parallel_0}|^2$ &  $(S_1 - S_3) - 2 S_3 \cos2\theta_\pi$
\\ \hline
$|{\cal A}^T_{\perp_1}|^2$ & $S^T_1 + \alpha_{\Lambda_c} S^T_2 \cos\theta_{\Lambda} \cos\theta_\pi + S^T_3 \cos2\theta_\pi$
\\ \hline
$|{\cal A}^T_{\parallel_1}|^2$ & $S^T_1 + \alpha_{\Lambda_c} S^T_2 \cos\theta_{\Lambda} \cos\theta_\pi + S^T_3 \cos2\theta_\pi$
\\ \hline
$|{\cal A}^T_{\perp_0}|^2$ & $(S_1^T - S_3^T) - 2 S_3^T \cos2\theta_\pi$
\\ \hline
$|{\cal A}^T_{\parallel_0}|^2$ & $(S_1^T - S_3^T) - 2 S_3^T \cos2\theta_\pi$
\\ \hline
\end{tabular}
\caption{\label{tab:NS} \small The enumeration of ${\cal N}^S_i |{\cal A}_i|^2$ pieces of Eq.~\eqref{eq:FFDW}.}
\end{center}
\end{table}
%%%%%%%%%%%%%%%%%%%%%%%%%%%%%%%%%%%%%%%%%%%%

%%%%%%%%%%%%%%%%%%%%%%%%%%%%%%%%%%%%%%%%%%%%
\begin{table}
{\scriptsize
\begin{center}
\begin{tabular}{|c|c|} 
\hline
Transversity Amplitudes & ${\cal N}^R$ 
\\ \hline
${\rm Re}[{\cal A}_{\perp_t} {\cal A}_{\parallel_t}^*]$ & $2 \alpha_{\Lambda_c} S_t \cos\theta_{\Lambda}$
\\ \hline
${\rm Re}[{\cal A}_{\perp_t} {\cal A}_{\parallel_1}^*]$ & $\alpha_{\Lambda_c} R_t \sin\theta_{\Lambda} \sin\theta_\pi \cos\phi_\pi$
\\ \hline
${\rm Re}[{\cal A}_{\perp_t} {\cal A}_{\perp_0}^*]$ & $-\sqrt{2} R_t \cos\theta_\pi$
\\ \hline
${\rm Re}[{\cal A}_{\perp_t} {\cal A}_{\parallel_0}^*]$ & $-\sqrt{2} \alpha_{\Lambda_c} R_t \cos\theta_{\Lambda} \cos\theta_\pi$
\\ \hline
${\rm Re}[{\cal A}_{\perp_t} {\cal A}^{T*}_{\parallel_1}]$ & $ \alpha_{\Lambda_c} R_t^T \sin\theta_{\Lambda} \sin\theta_\pi \cos\phi_\pi$
\\ \hline
${\rm Re}[{\cal A}_{\perp_t} {\cal A}^{T*}_{\perp_0}]$ & $ - \sqrt{2} R_t^T \cos\theta_\pi $
\\ \hline
${\rm Re}[{\cal A}_{\perp_t} {\cal A}^{T*}_{\parallel_0}]$ & $ - \sqrt{2} \alpha_{\Lambda_c} R_t^T \cos\theta_{\Lambda} \cos\theta_\pi $
\\ \hline
${\rm Re}[{\cal A}_{\parallel_t} {\cal A}_{\perp_1}^*]$ & $ - \alpha_{\Lambda_c} R_t \sin\theta_{\Lambda} \sin\theta_\pi \cos\phi_\pi$
\\ \hline
${\rm Re}[{\cal A}_{\parallel_t} {\cal A}_{\perp_0}^*]$ & $-\sqrt{2} \alpha_{\Lambda_c} R_t \cos\theta_{\Lambda} \cos\theta_\pi$
\\ \hline
${\rm Re}[{\cal A}_{\parallel_t} {\cal A}_{\parallel_0}^*]$ & $-\sqrt{2} R_t \cos\theta_\pi$
\\ \hline
${\rm Re}[{\cal A}_{\parallel_t} {\cal A}^{T*}_{\perp_1}]$ & $- \alpha_{\Lambda_c} R_t^T \sin\theta_{\Lambda} \sin\theta_\pi \cos\phi_\pi$
\\ \hline
${\rm Re}[{\cal A}_{\parallel_t} {\cal A}^{T*}_{\perp_0}]$ & $- \sqrt{2} \alpha_{\Lambda_c} R_t^T \cos\theta_{\Lambda} \cos\theta_\pi$
\\ \hline
${\rm Re}[{\cal A}_{\parallel_t} {\cal A}^{T*}_{\parallel_0}]$ & $- \sqrt{2} R_t^T  \cos\theta_\pi$
\\ \hline
${\rm Re}[{\cal A}_{\perp_1} {\cal A}_{\parallel_1}^*]$ & $2 S_2 \cos\theta_\pi +  2 \alpha_{\Lambda_c} \cos\theta_{\Lambda} (S_1 + S_3 \cos2\theta_\pi)$
\\ \hline
${\rm Re}[{\cal A}_{\perp_1} {\cal A}_{\perp_0}^*]$ & $\sqrt{2} \alpha_{\Lambda_c} S_2 \sin\theta_{\Lambda} \sin\theta_\pi \cos\phi_\pi$
\\ \hline
${\rm Re}[{\cal A}_{\perp_1} {\cal A}_{\parallel_0}^*]$ & $- 2\sqrt{2} \alpha_{\Lambda_c} S_3 \sin\theta_{\Lambda} \sin2\theta_\pi \cos\phi_\pi$
\\ \hline
${\rm Re}[{\cal A}_{\perp_1} {\cal A}^{T*}_{\perp_1}]$ & $ R_1 + \alpha_{\Lambda_c} R_2 \cos\theta_{\Lambda} \cos\theta_\pi + R_3 \cos2\theta_\pi$
\\ \hline
${\rm Re}[{\cal A}_{\perp_1} {\cal A}^{T*}_{\parallel_1}]$ & $R_2 \cos\theta_\pi + \alpha_{\Lambda_c} \cos\theta_{\Lambda} (R_1 + R_3 \cos2\theta_\pi)$
\\ \hline
${\rm Re}[{\cal A}_{\perp_1} {\cal A}^{T*}_{\perp_0}]$ & $(\alpha_{\Lambda_c} R_2 / \sqrt{2}) \sin\theta_{\Lambda} \sin\theta_\pi \cos\phi_\pi$
\\ \hline
${\rm Re}[{\cal A}_{\perp_1} {\cal A}^{T*}_{\parallel_0}]$ & $-\sqrt{2} \alpha_{\Lambda_c} R_3 \sin\theta_{\Lambda} \sin2\theta_\pi \cos\phi_\pi$
\\ \hline
${\rm Re}[{\cal A}_{\parallel_1} {\cal A}_{\perp_0}^*]$ & $2\sqrt{2} \alpha_{\Lambda_c} S_3 \sin\theta_{\Lambda} \sin2\theta_\pi \cos\phi_\pi$
\\ \hline
${\rm Re}[{\cal A}_{\parallel_1} {\cal A}_{\parallel_0}^*]$ & $-\sqrt{2} \alpha_{\Lambda_c} S_2 \sin\theta_{\Lambda} \sin\theta_\pi \cos\phi_\pi$
\\ \hline
${\rm Re}[{\cal A}_{\parallel_1} {\cal A}^{T*}_{\perp_1}]$ & $R_2 \cos\theta_\pi + \alpha_{\Lambda_c} \cos\theta_{\Lambda} (R_1 + R_3 \cos2\theta_\pi)$
\\ \hline
${\rm Re}[{\cal A}_{\parallel_1} {\cal A}^{T*}_{\parallel_1}]$ & $ R_1 + \alpha_{\Lambda_c} R_2 \cos\theta_{\Lambda} \cos\theta_\pi + R_3 \cos2\theta_\pi$
\\ \hline
${\rm Re}[{\cal A}_{\parallel_1} {\cal A}^{T*}_{\perp_0}]$ & $ \sqrt{2} \alpha_{\Lambda_c} R_3 \sin\theta_{\Lambda} \sin2\theta_\pi \cos\phi_\pi$
\\ \hline
${\rm Re}[{\cal A}_{\parallel_1} {\cal A}^{T*}_{\parallel_0}]$ & $-(\alpha_{\Lambda_c} R_2/\sqrt{2}) \sin\theta_{\Lambda} \sin\theta_\pi \cos\phi_\pi$
\\ \hline
${\rm Re}[{\cal A}_{\perp_0} {\cal A}_{\parallel_0}^*]$ & $2 \alpha_{\Lambda_c} \cos\theta_{\Lambda} (S_1 - S_3 - 2 S_3 \cos2\theta_\pi)$
\\ \hline
${\rm Re}[{\cal A}_{\perp_0} {\cal A}^{T*}_{\perp_1}]$ & $(\alpha_{\Lambda_c} R_2/\sqrt{2}) \sin\theta_{\Lambda} \sin\theta_\pi \cos\phi_\pi$
\\ \hline
${\rm Re}[{\cal A}_{\perp_0} {\cal A}^{T*}_{\parallel_1}]$ & $ \sqrt{2} \alpha_{\Lambda_c} R_3 \sin\theta_{\Lambda} \sin2\theta_\pi \cos\phi_\pi$
\\ \hline
${\rm Re}[{\cal A}_{\perp_0} {\cal A}^{T*}_{\perp_0}]$ & $ R_1-R_3 - 2 R_3 \cos2\theta_\pi$
\\ \hline
${\rm Re}[{\cal A}_{\perp_0} {\cal A}^{T*}_{\parallel_0}]$ & $ \alpha_{\Lambda_c} \cos\theta_{\Lambda} (R_1-R_3 - 2 R_3 \cos2\theta_\pi)$
\\ \hline
${\rm Re}[{\cal A}_{\parallel_0} {\cal A}^{T*}_{\perp_1}]$ & $ - \sqrt{2} \alpha_{\Lambda_c} R_3 \sin\theta_{\Lambda} \sin2\theta_\pi \cos\phi_\pi$
\\ \hline
${\rm Re}[{\cal A}_{\parallel_0} {\cal A}^{T*}_{\parallel_1}]$ & $- (\alpha_{\Lambda_c} R_2/\sqrt{2}) \sin\theta_{\Lambda} \sin\theta_\pi \cos\phi_\pi$
\\ \hline
${\rm Re}[{\cal A}_{\parallel_0} {\cal A}^{T*}_{\perp_0}]$ & $\alpha_{\Lambda_c} \cos\theta_{\Lambda} (R_1-R_3 - 2 R_3 \cos2\theta_\pi)$
\\ \hline
${\rm Re}[{\cal A}_{\parallel_0} {\cal A}^{T*}_{\parallel_0}]$ & $R_1-R_3 - 2 R_3 \cos2\theta_\pi$
\\ \hline
${\rm Re}[{\cal A}^T_{\perp_1} {\cal A}^{T*}_{\parallel_1}]$ & $2 S_2^T \cos\theta_\pi + 2 \alpha_{\Lambda_c} \cos\theta_{\Lambda} (S_1^T + S_3^T \cos2\theta_\pi)$
\\ \hline
${\rm Re}[{\cal A}^T_{\perp_1} {\cal A}^{T*}_{\perp_0}]$ & $\sqrt{2} \alpha_{\Lambda_c} S_2^T \sin\theta_{\Lambda} \sin\theta_\pi \cos\phi_\pi$
\\ \hline
${\rm Re}[{\cal A}^T_{\perp_1} {\cal A}^{T*}_{\parallel_0}]$ & $-2\sqrt{2} \alpha_{\Lambda_c} S_3^T \sin\theta_{\Lambda} \sin2\theta_\pi \cos\phi_\pi$
\\ \hline
${\rm Re}[{\cal A}^T_{\parallel_1} {\cal A}^{T*}_{\perp_0}]$ & $2\sqrt{2}\alpha_{\Lambda_c} S_3^T \sin\theta_{\Lambda} \sin2\theta_\pi \cos\phi_\pi$
\\ \hline
${\rm Re}[{\cal A}^T_{\parallel_1} {\cal A}^{T*}_{\parallel_0}]$ & $-\sqrt{2}\alpha_{\Lambda_c} S_2^T \sin\theta_{\Lambda} \sin\theta_\pi \cos\phi_\pi$
\\ \hline
${\rm Re}[{\cal A}^T_{\perp_0} {\cal A}^{T*}_{\parallel_0}]$ & $2 \alpha_{\Lambda_c} \cos\theta_{\Lambda} (S_1^T - S_3^T - 2 S_3^T \cos2\theta_\pi)$
\\ \hline
\end{tabular}
\caption{\label{tab:NR} \small The enumeration of ${\cal N}^R_{i,j} {\rm Re}[{\cal A}_i {\cal A}_j^*]$ pieces of Eq.~\eqref{eq:FFDW}.}
\end{center}
}
\end{table}
%%%%%%%%%%%%%%%%%%%%%%%%%%%%%%%%%%%%%%%%%%%%

%%%%%%%%%%%%%%%%%%%%%%%%%%%%%%%%%%%%%%%%%%%%
\begin{table}
\begin{center}
\begin{tabular}{|c|c|} 
\hline
Transversity Amplitudes & ${\cal N}^I$ 
\\ \hline
${\rm Im}[{\cal A}_{\perp_t} {\cal A}_{\perp_1}^*]$ & $- \alpha_{\Lambda_c} R_t \sin\theta_{\Lambda} \sin\theta_\pi \sin\phi_\pi$
\\ \hline
${\rm Im}[{\cal A}_{\perp_t} {\cal A}^{T*}_{\perp_1}]$ & $- \alpha_{\Lambda_c} R_t^T \sin\theta_{\Lambda} \sin\theta_\pi \sin\phi_\pi$
\\ \hline
${\rm Im}[{\cal A}_{\parallel_t} {\cal A}_{\parallel_1}^*]$ & $ \alpha_{\Lambda_c} R_t \sin\theta_{\Lambda} \sin\theta_\pi \sin\phi_\pi$
\\ \hline
${\rm Im}[{\cal A}_{\parallel_t} {\cal A}^{T*}_{\parallel_1}]$ & $ \alpha_{\Lambda_c} R_t^T \sin\theta_{\Lambda} \sin\theta_\pi \sin\phi_\pi$
\\ \hline
${\rm Im}[{\cal A}_{\perp_1} {\cal A}_{\perp_0}^*]$ & $2\sqrt{2} \alpha_{\Lambda_c} S_3 \sin\theta_{\Lambda} \sin2\theta_\pi \sin\phi_\pi$
\\ \hline
${\rm Im}[{\cal A}_{\perp_1} {\cal A}_{\parallel_0}^*]$ & $- \sqrt{2} \alpha_{\Lambda_c} S_2 \sin\theta_{\Lambda} \sin\theta_\pi \sin\phi_\pi$
\\ \hline
${\rm Im}[{\cal A}_{\perp_1} {\cal A}^{T*}_{\perp_0}]$ & $\sqrt{2}\alpha_{\Lambda_c} R_3 \sin\theta_{\Lambda} \sin2\theta_\pi \sin\phi_\pi$
\\ \hline
${\rm Im}[{\cal A}_{\perp_1} {\cal A}^{T*}_{\parallel_0}]$ & $- (\alpha_{\Lambda_c} R_2 / \sqrt{2}) \sin\theta_{\Lambda} \sin\theta_\pi \sin\phi_\pi$
\\ \hline
${\rm Im}[{\cal A}_{\parallel_1} {\cal A}_{\perp_0}^*]$ & $ \sqrt{2}\alpha_{\Lambda_c} S_2  \sin\theta_{\Lambda} \sin\theta_\pi \sin\phi_\pi$
\\ \hline
${\rm Im}[{\cal A}_{\parallel_1} {\cal A}_{\parallel_0}^*]$ & $-2\sqrt{2} \alpha_{\Lambda_c} S_3 \sin\theta_{\Lambda} \sin2\theta_\pi \sin\phi_\pi$
\\ \hline
${\rm Im}[{\cal A}_{\parallel_1} {\cal A}^{T*}_{\perp_0}]$ & $ (\alpha_{\Lambda_c} R_2/\sqrt{2}) \sin\theta_{\Lambda} \sin\theta_\pi \sin\phi_\pi$
\\ \hline
${\rm Im}[{\cal A}_{\parallel_1} {\cal A}^{T*}_{\parallel_0}]$ & $- \sqrt{2} \alpha_{\Lambda_c} R_3 \sin\theta_{\Lambda} \sin2\theta_\pi \sin\phi_\pi$
\\ \hline
${\rm Im}[{\cal A}_{\perp_0} {\cal A}^{T*}_{\perp_1}]$ & $ - \sqrt{2} \alpha_{\Lambda_c} R_3 \sin\theta_{\Lambda} \sin2\theta_\pi \sin\phi_\pi$
\\ \hline
${\rm Im}[{\cal A}_{\perp_0} {\cal A}^{T*}_{\parallel_1}]$ & $ - (\alpha_{\Lambda_c} R_2/\sqrt{2}) \sin\theta_{\Lambda} \sin\theta_\pi \sin\phi_\pi$
\\ \hline
${\rm Im}[{\cal A}_{\parallel_0} {\cal A}^{T*}_{\perp_1}]$ & $ (\alpha_{\Lambda_c} R_2/\sqrt{2}) \sin\theta_{\Lambda} \sin\theta_\pi \sin\phi_\pi$
\\ \hline
${\rm Im}[{\cal A}_{\parallel_0} {\cal A}^{T*}_{\parallel_1}]$ & $ \sqrt{2} \alpha_{\Lambda_c} R_3 \sin\theta_{\Lambda} \sin2\theta_\pi \sin\phi_\pi$
\\ \hline
${\rm Im}[{\cal A}^T_{\perp_1} {\cal A}^{T*}_{\perp_0}]$ & $2\sqrt{2} \alpha_{\Lambda_c} S_3^T \sin\theta_{\Lambda} \sin2\theta_\pi \sin\phi_\pi$
\\ \hline
${\rm Im}[{\cal A}^T_{\perp_1} {\cal A}^{T*}_{\parallel_0}]$ & $-\sqrt{2} \alpha_{\Lambda_c} S_2^T \sin\theta_{\Lambda} \sin\theta_\pi \sin\phi_\pi$
\\ \hline
${\rm Im}[{\cal A}^T_{\parallel_1} {\cal A}^{T*}_{\perp_0}]$ & $\sqrt{2}\alpha_{\Lambda_c} S_2^T \sin\theta_{\Lambda} \sin\theta_\pi \sin\phi_\pi$
\\ \hline
${\rm Im}[{\cal A}^T_{\parallel_1} {\cal A}^{T*}_{\parallel_0}]$ & $-2\sqrt{2}\alpha_{\Lambda_c} S_3^T \sin\theta_{\Lambda} \sin2\theta_\pi \sin\phi_\pi$
\\ \hline
\end{tabular}
\caption{\label{tab:NI} \small The enumeration of ${\cal N}^I_{i,j} {\rm Im}[{\cal A}_i {\cal A}_j^*]$ pieces of Eq.~\eqref{eq:FFDW}.}
\end{center}
\end{table}
%%%%%%%%%%%%%%%%%%%%%%%%%%%%%%%%%%%%%%%%%%%%

To make the expressions more compact, we define the following dimensionless parameters
\begin{equation}
\kappa_\tau \equiv  \frac{m_\tau}{\sqrt{q^2}},
\quad 
\kappa_\pi \equiv  \frac{m_\pi}{\sqrt{q^2}},
\quad
\omega_\pi \equiv  \frac{E_\pi}{\sqrt{q^2}}.
\end{equation}
The dimensionless factors in Table~\ref{tab:NS}, \ref{tab:NR}, and \ref{tab:NI} are given by
\begin{align}
&S_t = 2 \omega _{\pi } \kappa _{\tau }^2-\kappa _{\tau }^4-\kappa _{\pi }^2,
\label{eq:St}
\\[2mm]
&S_1 = \frac{\kappa _{\tau }^2 }{8 \left(\omega _{\pi
	}^2 - \kappa _{\pi }^2\right)}\Big[\kappa _{\pi }^2 \left(-6 \omega _{\pi } \kappa
_{\tau }^2+3 \kappa _{\tau }^4+4 \omega _{\pi }^2+10 \omega _{\pi
}-5\right)
\nonumber\\[2mm]
&\qquad + \left( 2 \omega _{\pi } - \kappa_{\tau}^2\right)  \left(2 \omega _{\pi }^2+2 \omega _{\pi
}-1\right) \kappa _{\tau }^2-3 \kappa _{\pi }^4+6 \left(1-2 \omega _{\pi
}\right) \omega _{\pi }^2\Big],
\\[2mm]
&S_2 = \frac{\kappa _{\tau }^2 \left(\kappa _{\pi }^2-2 \omega _{\pi }+1\right)
	 \left(\omega _{\pi }-\kappa _{\tau
	}^2\right)}{\sqrt{\omega _{\pi }^2-\kappa _{\pi }^2}},
\\[2mm]
&S_3 = \frac{\kappa _{\tau }^2}{8 \left(\omega _{\pi }^2 - \kappa _{\pi }^2\right)} \Big[\kappa _{\pi }^2 \left(-2 \omega _{\pi } \kappa
_{\tau }^2+\kappa _{\tau }^4+4 \omega _{\pi }^2-2 \omega _{\pi
}+1\right)
\nonumber\\[2mm]
&\qquad + \left( \kappa _{\tau
}^2-2 \omega _{\pi }\right)  \left(2 \omega _{\pi }^2-6 \omega _{\pi }+3\right)
\kappa _{\tau }^2-\kappa _{\pi }^4+2 \left(1-2 \omega _{\pi }\right) \omega
_{\pi }^2\Big],
\\[2mm]
&S^T_1 = \frac{1}{2 \left(\kappa
	_{\pi }^2-\omega _{\pi }^2\right)} \Big\{ \kappa _{\pi }^4 \left(2 \omega _{\pi } \kappa _{\tau }^2+5 \kappa
_{\tau }^4+2 \omega _{\pi }-3\right) +4
\omega _{\pi }^2 \kappa _{\tau }^2 \left[\left(3 \omega _{\pi }-1\right)
\kappa _{\tau }^2-\omega _{\pi }\right]
\nonumber\\[2mm]
&\qquad +\kappa _{\pi }^2 \left[\left(-6 \omega
_{\pi }^2-10 \omega _{\pi }+3\right) \kappa _{\tau }^4+2 \left(3-2 \omega
_{\pi }\right) \omega _{\pi } \kappa _{\tau }^2+2 \omega _{\pi }^2\right]-\kappa _{\pi }^6 \Big\},
\\[2mm]
& S^T_2 = \frac{4 \kappa _{\tau }^2 \left(\kappa _{\pi }^2-2 \omega _{\pi }+1\right)
	 \left(\kappa _{\pi }^2-\omega _{\pi } \kappa _{\tau
	}^2\right)}{\sqrt{\omega _{\pi }^2-\kappa _{\pi }^2}},
\\[2mm]
& S^T_3 = \frac{1}{2
	\left(\omega _{\pi }^2 - \kappa _{\pi }^2\right)} \Big\{ \kappa _{\pi }^4 \left(-6 \omega _{\pi } \kappa _{\tau }^2+\kappa
_{\tau }^4-6 \omega _{\pi }+1\right) -4 \omega _{\pi }^2 \kappa _{\tau }^2 \left[\left(\omega _{\pi
}-1\right) \kappa _{\tau }^2+\omega _{\pi }\right]
\nonumber\\[2mm]
& \qquad +\kappa _{\pi }^2 \left[\left(2 \omega
_{\pi }^2-2 \omega _{\pi }-1\right) \kappa _{\tau }^4+2 \omega _{\pi }
\left(6 \omega _{\pi }-1\right) \kappa _{\tau }^2+2 \omega _{\pi
}^2\right] +3 \kappa _{\pi }^6 \Big\},
\end{align}
and
\begin{align}
& R_t = \frac{\sqrt{2} \left(\omega _{\pi }-1\right) \kappa _{\tau } \left(2 \omega _{\pi } \kappa _{\tau }^2 - \kappa _{\tau }^4 - \kappa_{\pi }^2\right)}{\sqrt{\omega _{\pi }^2-\kappa _{\pi }^2}},
\\[2mm]
& R_t^T = \frac{2 \sqrt{2} \left[\kappa _{\pi }^2 \left(-2 \omega _{\pi } \kappa _{\tau
	}^2+\kappa _{\tau }^4-\omega _{\pi }\right)-\omega _{\pi } \kappa _{\tau
	}^2 \left(\kappa _{\tau }^2-2 \omega _{\pi }\right)+\kappa _{\pi
	}^4\right]}{\sqrt{\omega _{\pi }^2-\kappa _{\pi }^2}},
\\[2mm]
&R_1 = \frac{\kappa _{\tau }}{2 \left(\kappa _{\pi }^2-\omega _{\pi }^2\right)} \Big\{\kappa_{\pi }^2 \left[\left(\omega _{\pi }+2\right) \kappa _{\tau }^4+\left(4 \omega _{\pi }^2+8 \omega
_{\pi }-6\right) \kappa _{\tau }^2-4 \omega _{\pi }^2+\omega _{\pi }\right] 
\nonumber\\[2mm]
&\qquad +\kappa _{\pi }^4 \left(-6 \kappa _{\tau }^2+\omega _{\pi }+2\right) +\omega _{\pi } \kappa
_{\tau }^2 \left[\left(1-4 \omega _{\pi }\right) \kappa _{\tau }^2-4 \left(\omega _{\pi }-1\right)
\omega _{\pi }\right]\Big\},
\\[2mm]
&R_2 = \frac{2 \kappa _{\tau } \left(\kappa _{\tau }^4-\kappa _{\pi }^2\right) \left(\kappa _{\pi }^2-2 \omega
	_{\pi }+1\right)}{\sqrt{\omega _{\pi }^2-\kappa _{\pi }^2}},
\\[2mm]
&R_3 = \frac{\kappa _{\tau } }{2 \left(\kappa _{\pi }^2-\omega _{\pi }^2\right)} \Big\{\kappa
_{\pi }^2 \left[\left(3 \omega _{\pi }-2\right) \kappa _{\tau }^4+\left(-4 \omega _{\pi }^2+8 \omega
_{\pi }-2\right) \kappa _{\tau }^2+\left(3-4 \omega _{\pi }\right) \omega _{\pi }\right]
\nonumber\\[2mm] 
&\qquad +\kappa _{\pi }^4 \left(-2 \kappa _{\tau }^2+3 \omega _{\pi }-2\right)+\omega _{\pi
} \kappa _{\tau }^2 \left[\left(3-4 \omega _{\pi }\right) \kappa _{\tau }^2+4 \left(\omega _{\pi
}-1\right) \omega _{\pi }\right]\Big\}.
\label{eq:R3}
\end{align}

\bibliographystyle{JHEP}
\bibliography{ref}

\end{document}